# Ordered Sets for Data Analysis

Part 1: Relations, Orders, Lattices, Formal Concept Analysis

Sergei O. Kuznetsov

National Research University Higher School of Economics

Moscow, Russia, 2019

# Contents

# Preface

Most data of the early days of data analysis were nominal or numerical: distances, lengths, heights, widths, etc. were categorized or measured and provided input for the analysis.

Numerical data were mostly considered in terms of vector spaces, where dependencies are naturally described in terms of regressions, compact regions, and separating surfaces.

Today we encounter problems where we have to analyze much more complex data, such as texts, chemical structures, biological structures, fingerprints, images, streams, and so on. Often it is not easy at all to construct numerical or binary "features" for these descriptions that would allow to embed transformed descriptions in vector space in order to obtain reasonable results: the data "resist" and demand treating them differently, with due respect to their *structure*.

Symbolic data are usually given by descriptions that can be ordered with respect to generality relation, where two descriptions are comparable if one of them is "more general" than the second one, meaning that the former "covers" more objects than the former. It is hard to imagine data that do not induce a natural relation of this kind. This relation is naturally transitive: if a description $A$ is more general than description $B$, while $B$ is more general than description $C$, then $A$ is more general than description $C$. This kind of generality goes hand in hand with amount of details, "precision", of descriptions, which is often related to some natural size of description. The less detailed description we have, the more general it is, matching (covering) more objects. The relationship between thus understood precision and generality is naturally captured by the mathematical notion of Galois connection.

This book dwells on mathematical and algorithmic issues of data analysis based on generality order of descriptions and respective precision...

To speak of these topics correctly, we have to go some way getting acquainted with the important notions of relation and order theory. On the one hand, data often have complex structure with natural order on it. On the other hand, many symbolic methods of data analysis and machine learning allow to compare the obtained classifiers w.r.t. their generality, which is also an order relation. Efficient algorithms are very important in data analysis, especially when one deals with big data, so scalability is a real issue. That is why we analyze computational complexity of algorithms and problems of data analysis. We start from the basic definitions and facts of algorithmic complexity theory and analyze complexity of various tools of data analysis we consider. The tools and methods of data analysis, like computing taxonomies, groups of similar objects (concepts and n-clusters), dependencies in data, classification, etc., are illustrated with applications in particular subject domains, from chemoinformatics to text mining and natural language processing.

The structure of the book is as follows:

1. Relations, binary relations, their matrices and graphs. Operations over relations, their properties, and types of relations.

   **Main topics**: relation,

2. Partial order order and its diagram

   **Main topics**: graphs and diagrams of partial order, topological sorting, Galois connection between two partial orders

3. Introduction to the theory of algorithmic complexity

   **Main topics**: Worst-case complexity, classes NP, co-NP, #P. Delay, total complexity, amortized complexity.

4. Semilattices and lattices

   **Main topics**: infimums, supremums, classes of lattices

5. Introduction to Formal Concept Analysis (FCA)

   **Main topics**: (formal) concept, concept extent, concept intent, order on concepts, concept lattice

6. Implications and functional dependencies

   **Main topics**: implication bases, implicational closure, Attribute Exploration

7. Association rules

   **Main topics**: confidence, support of association rules, concept lattice as a concise representation of association rules

8. Algorithmic problems of computing concept lattices and implication bases

   **Main topics**: intractability of generation of concept lattices and implication bases, efficient algorithms

# 1 Relations, their matrices and graphs

In this section we consider main definitions and facts about binary relations, their matrix and graph representations, operations on relations, and properties of relations. We assume that the reader is familiar with basic notions of Boolean algebras and first-order logic and set theory.

In life we know many relations between entities we encounter, they are usually informally defined through their "meaning" or *in*tentionally. In mathematics, relation is defined *ex*tentionally, through the set of finite tuples of objects that are in the given relation.

**Definition 1.1.** Cartesian (direct) product of sets $A_1, \ldots, A_n$ *is a set of tuples (finite sequences) of size $n$ where $i$th element belongs to $A_i$:*

$$A_1 \times \ldots \times A_n := \{(a_1, \ldots, a_n) \mid a_1 \in A_1, \ldots a_n \in A_n\}.$$

If $A_i = A$ for all $i \in \{1, \ldots, n\}$, one speaks about $n$th power of set $A$.
In particular,

**Definition 1.2.** Cartesian (direct) product of sets $A$ and $B$ *is a set of pairs, so that the first element of each pair belongs to (is an element of) $A$ and the second element belongs to $B$:*

$$A \times B := \{(a, b) \mid a \in A, b \in B\}.$$

If sets $A$ and $B$ coincide, we have a Cartesian square of $A$ $(B)$.

**Definition 1.3.** $n$-ary relation $R$ on sets $A_1, \ldots, A_n$ *is a subset of Cartesian product of sets $A_1, \ldots, A_n$: $R \subseteq A_1 \times \ldots \times A_n$.*

**Example 1.1.** 11-nary relation "to play in one man football team at time T" is given as a subset of 11-degree of the set of men.

In what follows we will mostly deal with binary relations, i.e., $n$-ary relations where $n = 2$.

**Definition 1.4.** Binary relation $R$ between sets $A$ and $B$ *is a subset of Cartesian product of sets $A$ and $B$:*

$$R \subseteq A \times B$$

.

Example. In data analysis a typical binary relation is defined for a set of objects and a set of their possible attributes. For example, objects are transactions in a supermarket, attributes are items in the supermarket, and the relation consists of pairs (transaction, item) such that the item occurs in the transaction.

If $A = B$, then one says that $R$ is a *relation on set $A$.*

Examples. Consider $F$ to be a set of females and $M$ to be a set of males, then the relation "to be a mother of" is given on the Cartesian product $F \times (F \cup M)$. The relation "to be a parent" is naturally given on the Cartesian product $(F \cup M) \times (F \cup M)$. In mathematics very basic relations on the set of natural numbers are those of "equality," "be less than or equal," "be smaller," etc.

Any binary relation on a pair of sets $A$ and $B$ can be represented as a relation on one set $A \cup B$. Hence, without loss of generality one can speak about relation on one set. However, if we want to stress that $A$ and $B$ have absolutely different nature, it is better to define the relation on two sets $A$ and $B$.

In what follows we will mostly speak of binary relations, so we will use the term "relation" meaning a binary relation and use the term "n-ary relation" for relations with $n > 2$.

**Definition 1.5.** Infix form *for denoting relation $R$ is notation of the form $aRb \Leftrightarrow (a, b) \in R \subseteq A \times B$.*

This form is used, e.g., for writing down usual binary relations on numbers, such as "smaller than," etc.

Important cases of binary relations are *identity* $I := \{(a, a) \mid a \in A\}$ and *universal* $U := \{(a, b) \mid a \in A, b \in A)\}$.

## 1.1 Properties of binary relations

Here we give a list of main properties of binary relations which are most important from the viewpoint of applications in data science:

**Definition 1.6.** *Relation $R \subseteq A \times A$ is called*

reflexive *if $\forall a \in A \ aRa$*

antireflexive *if $\forall a \in A \quad \neg(aRa) \ (\Leftrightarrow aR^c a)$*

symmetric *if $\forall a, b \in A \quad aRb \Rightarrow bRa$*

assymmetric *if $\forall a, b \in A \quad aRb \Rightarrow \neg(bRa) \ (\Leftrightarrow bR^c a)$*

antisymmetric *if $\forall a, b \in A \quad aRb \ \& \ bRa \Rightarrow a = b$*

transitive *if $\forall a, b, c \in A \quad aRb \ \& \ bRc \Rightarrow aRc$*

linear or complete *if $\forall a, b \in A \quad a \neq b \Rightarrow aRb \vee bRa$.*

Note that nonreflexive relation is not always antireflexive and nonsymmetric relations is not always asymmetric. Below we consider several important types of binary relations given by combinations of properties from this list.

## 1.2 Matrix and graph representation of relations

Two natural representations of binary relations are binary matrices and graphs. The properties of binary relations listed above are easily interpreted in their terms.

*Matrix (datatable) of binary relation $R \subseteq A \times A$* looks as follows:

$$
\begin{array}{c|ccc}
R & \cdots & a_j & \cdots \\
\hline
\vdots & & \vdots & \\
a_i & \cdots & \varepsilon_{ij} & \\
\vdots & & &
\end{array}
\qquad
\varepsilon_{ij} =
\begin{cases}
1, & \text{если } (a_i, a_j) \in R \\
0, & \text{если } (a_i, a_j) \notin R
\end{cases}
$$

From the viewpoint of the relational binary matrix the relation consists of pairs (**row**, **column**) so that the matrix entry defined by **row** and **column** is unit.

Graphs give another representation of binary relations in terms of drawings consisting of points (vertices) and lines (arcs or edges) connecting the points.

**Definition 1.7.** Directed graphs (digraph) *$G$ is a pair of the form $(V, A)$, where $V$ is a set of* vertices *of the graph and $A \subseteq V \times V$ is called the set of* arcs *of graph $G$.*

Directed graph on a set of vertices $V$ gives a binary relation on set $V$. From graph-theoretical viewpoint, the matrix defining the relation is the adjacency matrix of the relation graph.

Examples are directed graphs of parent relation on the set of humans, preference relation on some goods or commodities, "go before" relation on time instances, "more precise" relation on data descriptions.

**Definition 1.8.** (Undirected) graph *is a pair* $G = (V, E)$, *where* $V$ *is the set of* vertices *of the graph. The set* $E = \{\{v, u\} \mid v, u \in V\} \cup E_0$, *where* $E = \{\{v, u\} \mid v, u \in V\}$, *a set of* Unordered *pairs of elements of the set* $V$, *is called the set of* edges, *and* $E_0 \subseteq V$ *is the set of* loops. *If* $E_0 = \emptyset$, *then* $G$ *is called* loopless.

An undirected graph on set of vertices $V$ represent a symmetric relation on $V$, i.e. $R \subseteq V \times V$: $(a, b) \in R \Leftrightarrow (b, a) \in R$.

Examples of symmetric relations on sets of different nature are equality, similarity, sharing certain properties.

A relation on Cartesian product of two different sets is naturally represented by a directed bipartite graph.

**Definition 1.9.** Directed bipartite gaph *is a pair of the form* $(V, A)$, *where* $V = V_1 \cup V_2$, $V_1 \cap V_2 = \emptyset$, *and* $A \subseteq V_1 \times V_2$, *i.e., any arc from* $A$ *connects a vertex from* $V_1$ *with a vertex from* $V_2$. *Sets* $V_1$ *and* $V_2$ *are called* parts *of the graph.*

Examples are the relation "teaches" on the sets of teachers and pupils in a school, "matches" on sets of classifiers and object descriptions.

Symmetric relations on two different sets are naturally represented by undirected bipartite graphs.

**Definition 1.10.** Undirected bipartite graph *is a pair of the form* $(V, E)$, *where* $V$ *is a set of vertices,* $V = V_1 \cup V_2$ $V_1 \cap V_2 = \emptyset$, *and* $E = \{\{v_1, v_2\} \mid v_1 \in V_1, v_2 \in V_2\}$ *is a set of undirected edges connecting vertices from* $V_1$ *with vertices from* $V_2$.

An undirected bipartite graph naturally represents a relation on two sets, where direction of the relation is fixed, e.g., objects from set $O$ having attributes from set $A$. One does not need to draw a directed arc in this case, a (nondirected) edge is sufficient, since by default objects "have attributes", not other objects. Other examples are relations "to be married" on sets of men and women, "to be in communication" on human and governmental bodies.

So, a (directed) graph $G = (V, E)$ can be represented by *adjacency matrix* , this matrix being at the same time the matrix corresponding to the relation of the graph.

$$
\begin{array}{c|ccc}
 & \cdots & v_j & \cdots \\
\hline
\vdots & & \vdots & \\
v_i & \cdots & \varepsilon_{ij} & \\
\vdots & & &
\end{array}
\qquad
\varepsilon_{ij} = \begin{cases} 1, \ \text{если } (v_i, v_j) \in E \\ 0, \ \text{если } (v_i, v_j) \notin E \end{cases}
$$

In an undirected graph $\varepsilon_{ij} = \varepsilon_{ji}$.

**Definition 1.11.** *One says that vertex $v_i$ is* incident *to arc (edge) $e_j$, if* $\varepsilon_{ij} \neq 0$.

Any undirected graph $G = (V, E)$ can also be represented by *incidence matrix*:

$$
\begin{array}{c|ccc}
 & \cdots & e_j & \cdots \\
\hline
\vdots & & \vdots & \\
v_i & \cdots & \varepsilon_{ij} & \\
\vdots & & &
\end{array}
\qquad
\varepsilon_{ij} = \begin{cases} -1, \ \text{если } \exists v_k \in V \colon e_j = (v_i, v_k) \\ 1, \ \text{если } \exists v_k \in V \colon e_j = (v_k, v_i) \\ 0, \ \text{если } v_i \notin e_j \end{cases}
$$

For undirected graph

$$
\varepsilon_{ij} = \begin{cases} 1, \ \text{если } v_i \in e_j \\ 0, \ \text{если } v_i \notin e_j \end{cases}
$$

## 1.3 Subgraphs

**Definition 1.12.** *Graph $H = (V_H, E_H)$ is a* subgraph *of graph $G = (V_G, E_G)$ if all vertices and edges of $H$ are vertices and edges of $G$, i.e. $V_H \subseteq V_G$ and $E_H \subseteq E_G$. If $H$ is a subgraph of $G$, then $G$ is called a* supergraph *of $H$.*

In the language of relations a subgraph corresponds to a subset of relation of the graph.

**Definition 1.13.** *Graph $H = (V_H, E_H)$ is an* induced subgraph *of graph $G = (V_G, E_G)$ if $H$ is a subgraph $G$, and edges of $H$ are all edges of $G$ with both vertices lying in $H$.*

Graph $\Gamma_1 = (V_1, E_1)$ is isomorphic to graph $\Gamma_2 = (V_2, E_2)$, if there is a one-to-one mapping (bijection) $f$ from set vertices $V_1$ to set of vertices $V_2$ that respects edges, i.e., $(v_1, w_1) \in E_1 \leftrightarrow (f(v_1), f(w_1)) \in E_2$.

Isomorphic graphs can be given by same drawings consisting of vertices and arcs. For isomorphic graphs there is a bijection not only on vertices, but on the arcs (edges), so a bijection between respective relations.

Graph $\Gamma_1$ is isomorphic to a subgraph of graph $\Gamma_2$ if there a part of graph $\Gamma_2$ which is isomorphic to graph $\Gamma_1$.

## 1.4 Important graphs and subgraphs

**Definition 1.14.** Route (path) *is an alternating sequence of vertices and arcs (edges go after vertices) of the form $v_0, e_1, v_1, e_2, v_2, \ldots, e_k, v_k$, where two consecutive edges have a common vertex.*

*Chain* is a path where all edges are equal. *Simple chain* is a chain without repeating vertices. If two vertices are connected by a simple chain, they are called *connected*.

**Definition 1.15.** *A graph is called* connected *if all its vertices are pairwise connected.*

**Definition 1.16.** Connected component of a graph *is inclusion maximal subset of graph vertices, each pair of which are pairwise connected.*

Consider an example of an undirected graph w.r.t. the relation "be friends in a social network," then a connected component of this graph can be interpreted as a community.

**Definition 1.17.** (Directed) route *or* (directed) path *is an alternating sequence of graph vertices and arcs of the form $v_0, a_1, v_1, a_2, v_2, \ldots, a_k, v_k$, where the end of every arc (except for the last one) coincides with the beginning of the next arc, i.e., $a_i = (v_{i-1}, v_i)$.*

A *chain* is a path where all arcs are different. A *simple chain* is a chain without repeating vertices.

**Definition 1.18.** *Vertex $v_j$ is* reachable *from vertex $v_i$ if there exists a path starting in $v_i$ with the end in $v_j$ (it is considered that the vertex is reachable from itself).*

**Definition 1.19.** *A graph is* strongly connected *if every two vertices are reachable from each other. A graph is* one-way connected *if for every pair of vertices there is reachability in at least one direction.*

### 1.4.1 Cycles

**Definition 1.20.** *A* (directed) cycle (circuit) *is a (directed) path where the first and the last vertices coincide. A cycle where all vertices are different, except the first and the last ones, is called* simple.

**Definition 1.21.** *An* acyclic graph *is a directed graph with no directed cycles.*

**Definition 1.22.** *A* (directed) tree *is a (directed) graph without cycles.*

**Definition 1.23.** *A* rooted tree *is a tree with a designated vertex called* **root**.

In a rooted tree one has a designated direction from the root to leaves, therefore every rooted tree can be considered as directed.

**Definition 1.24.** *A star* is a tree of the form $G = (V, E)$, where $V = \{v_0\} \cup V_1$, $E = \{\{v_0, v_i\} \mid v_i \in V_1\}$.

**Definition 1.25.** *A (an undirected)* complete graph *is a (an undirected) graph $G = (V, E)$ where each pair of vertices is connected by an edge, i.e. $E = V \times V$.*

A complete graph on a set of vertices $V$ represents the universal relation on set $V$.

Example. A complete subgraph of the graph where vertices are people and edges correspond to the relation "be friends in a social network" can be interpreted as a strongly consolidated social group. A subset of a collection of texts where every two texts are very similar to each other can be considered as a cluster of similar texts.

A complete subgraph $G$ is called a *clique* if it cannot be extended by including one more adjacent vertex, i.e., any supergraph of $G$ is not complete. Some authors define cliques as complete graphs, calling cliques defined above maximal cliques.

## 1.5   Operations over Relations

Let $P$ and $R$ be two binary relations on set $A$. Since relations are sets, usual set-theoretic operations. Besides them one can apply other operations obtaining the following derived relations:

- *inverse* relation

  $$R^{-1} = R^d := \{(a, b) \mid (b, a) \in R\}$$

- *complement* of relation

  $$R^c = \overline{R} := \{(a, b) \mid (a, b) \notin R\}$$

- *incomparability* relation

  $$I_R = A \times A \setminus (R \cup R^d) = (R \cup R^d)^c = R^c \cap R^{cd}$$

- *product* of relations

  $$P \cdot R = \{(x, y) \mid \exists z \quad (x, z) \in P, (z, y) \in R\}$$

- *degree* of relation

  $$R^n = \underbrace{R \cdot R \cdot \ldots \cdot R}_{n \text{ раз}}$$

- *transitive closure* of relation

  $$R^T = R \cup R^2 \cup R^3 \cup \ldots = \bigcup_{i=1}^{\infty} R^i$$

- *reflexive transitive closure* of relation

  $$R^{TR} = R^T \cup I, \text{ where } I \text{ is identity, } I = \{(a, a) \mid a \in A\}.$$

- *kernel* of relation $R_k = R \cdot R^{-1}$.

## 1.6  Functions

An important type of binary relation is a function (mapping).

**Definition 1.26.** *Relation* $f \subseteq A \times B$ *is a* function (map, mapping) *from A to B (denoted by* $f : A \to B$*) if for every* $a \in A$ *There is* $b \in B$ *such that* $(a, b) \in f$ *and* $(a, b) \in f, (a, c) \in f \Rightarrow b = c.$

Function $f : A \to B$ is called

*injection (mapping in)* if $b = f(a_1)$ and $b = f(a_2) \Rightarrow a_1 = a_2$;

*surjection (mapping onto)* if for every $b \in B$ There is $a \in A$ such that $b = f(a)$ (or $\forall b \in B \; \exists a \in A \; b = f(a)$);

*bijection* if it is injection and surjection.

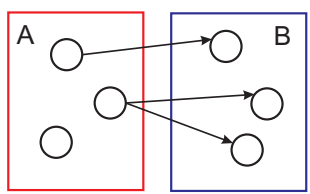

relation, but not a function

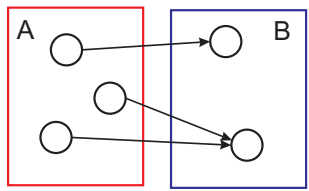

surjection, but not an injection

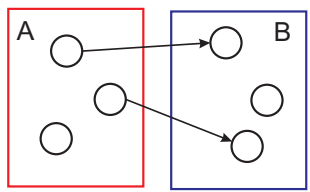

injection, but not a surjection

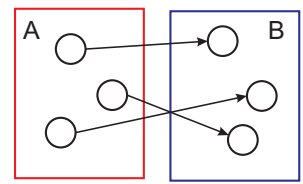

bijection

## 1.7  Types of binary relations

By combining properties from the list above one can obtain various types of binary relations. Most important relation types, both in mathematics and data science, are the following.

A binary relation on set $V$ is a *tolerance* if it reflexive and symmetric.

Similarity relation on objects is usually a tolerance. In applications it is often the case that similarity is defined through some similarity measure.

For example, in computational linguistics a text can be represented by the set of its key terms, and similarity of two texts given by sets $A$ and $B$ is defined by means of Jaccard measure:

$$\mathbf{Jac}(A, B) = \frac{|A \cap B|}{|A \cup B|}$$

Let $A$ and $B$ be called $\theta$-similar if $\mathrm{Jac}(A, B) \geq \theta$. It easy to see that the relation of $\theta$-similarity is reflexive, symmetric, but not always transitive.

**Definition 1.27.** *Consider a set $M$ and a tolerance relation on $M \times M$. Class of tolerance is a maximal subset of elements of $M$ where all pairs of elements belong to the relation.*

Since tolerance is not transitive in general, classes of tolerance are not necessarily disjoint. Tolerance is naturally represented by an undirected graph with loops. Since all vertices have loops, a usual convention is not to draw any loop at all. Classes of tolerance correspond to cliques in the graph of the relation.

In mathematics and applications one often encounters a special type of tolerance relation called equivalence.

1. *Equivalence* is a reflexive, symmetric, and transitive binary relation.

   There are numerous examples of equivalence in mathematics: equality relation, equality of natural numbers modulo some natural number, reachability in undirected graph, etc. In data science applications there are numerous examples of equivalence, such as "same description under certain data representation," e.g., equivalence relation on objects having same set of binary attributes.

   **Definition 1.28.** Class of equivalence *is a subset of elements of set $M$ that are equivalent to some element $x \in M$.*

   In contrast to tolerance classes, equivalence classes are disjoint, i.e., $A \cap B = \emptyset$ for any two equivalence classes $A$ and $AB$. The graph of equivalence relation consists of some (may be one) connected component(s), each of which corresponds to a class of equivalence and is a clique.

**Definition 1.29.** Partition *of set $M$ is a set of subsets $\{M_1, \ldots, M_n\}$, such that*

$$M_i \subseteq M, \bigcup_{i \in [1,n]} M_i = M, \quad \forall i, j \in [1,n] M_i \cap M_j = \emptyset.$$

There is a bijection between partition and equivalences on set $M$: Each equivalence relation on $M$ defines a partition (into classes of equivalence), and each partition of $M$ defines an equivalence relation "to belong to the same equivalence class".

*Quasiorder* or *preorder* is reflexive and transitive.

Examples. The relation "to be isomorphic to a subgraph of" is a quasiorder on sets of graphs. The relation of semantic entailment is a quasiorder on the set of logical formulas. Consider an object-attribute binary relation, where each object is described by a subset of attributes. The inclusion on subsets of attributes induces a quasiorder relation on objects. Two objects may have same descriptions, i.e. subsets of attributes within a fixed set of attributes, however they are still different objects.

2. *Partial order* is a reflexive, transitive, and antisymmetric binary relation;

   Examples. The relation "to be less than or equal" on real numbers, the relation "to be a subset of" on subsets of a set are typical examples of partial orders in mathematics.

3. *Strict partial order* is an antireflexive transitive binary relation.

   Strict order relation can be obtained from partial order by deleting pairs of equal elements.

   Examples. Relation "smaller than" on real numbers, "proper subset" on sets.

# Exercises

1. Show that Cartesian product is associative.

2. Let $F$ and $M$ be relations "to be father of" and "to be mother of" on the set of humans $H$, respectively. Using operations on relations construct relations "to be ancestor," "to be great grandfather", "to be a cousin".

3. Show that the product of relations is associative.

4. Show that the kernel relation is tolerance.

5. Show that a relation is tolerance if it is the kernel for a relation.

6. Show that for functions the respective kernel relation is equivalence.

7. Show that $R$ is equivalence iff $I \subseteq R$ and $R \cdot R^{-1} \subseteq R$.

8. Let $R$ be a relation on $A$. Show that

   $R$ is reflexive $\Leftrightarrow I \subseteq R$

   $R$ is antireflexive $\Leftrightarrow R \cap I = \emptyset$

   $R$ is symmetric $\Leftrightarrow R = R^{-1}$

   $R$ is asymmetric $\Leftrightarrow R \neq R^{-1}$

   $R$ is antisymmetric $\Leftrightarrow R \cap R^{-1} \subseteq I$

   $R$ is transitive $\Leftrightarrow R \cdot R \subseteq R$

   $R$ is linear $\Leftrightarrow R \cup I \cup R^{-1} = U$

9. Show that relation $R \subseteq A \times B$ is a bijection from $A$ to $B$ iff $R \cdot R^{-1} = R^{-1} \cdot R = I$.

10. Show that intersection of two equivalence relations is an equivalence relation.

11. Show that the product of two equivalence relations $R$ and $S$ is an equivalence relation iff $R \cdot S = S \cdot R$.

12. Show that the product $R \cdot S$ of two equivalence relations $R$ and $S$ contains $R$ and $S$, and is contained in any equivalence relation $T$ that contains $R$ and $S$.

13. Show that the union of quasiorder $Q$ with the inverse relation $Q^{-1}$ is an equivalence relation.

14. Compute the number of all injections from a finite set $A$ to a finite set $B$ for the case where $|A| < |B|$.

15. Compute the number of (non-induced) subgraphs of a finite graph with the set of edges $E$.

16. Degree of a vertex of a nondirected graph is the number of edges incident to the vertex. Prove that in an arbitrary graph the number of vertices with odd degree is even.

# 2   Introduction to algorithmic complexity

In this section we present some basic notions from the theory of algorithmic complexity that we will use below. For advanced introduction in the field consult standard references such as [13].

**Asymptotic notation $\Theta, O, \Omega, o$ and $\omega$.**

A software realization of any algorithm requires computational resources, which are measured in time and memory (space) of computation. Time and memory can be expressed in financial terms in various ways depending on technical and economical conditions.

Since in practice time and space required by an algorithm depend heavily on various "engineering" details, algorithmic complexity theory studies asymptotical efficiency of algorithms, which is independent of software realization and hardware used. This means that the theory studies the behavior of an algorithm in the limit when the size of input tends to infinity. Usually, an algorithm that is asymptotically more efficient than another one is more efficient for all data except for some amount of "small data." First we introduce several forms of standard "asymptotic notation," one of which is well-known $O$-notation.

The notation used in the description of asymptotic behavior of the algorithm time complexity uses functions with the range being the set of nonnegative integer numbers $N = \{0, 1, 2, \ldots\}$.

This notation is easily generalized to the domain of real numbers or constrained to subsets of the set of natural numbers.

### $\Theta$-notation

For a function $g(n) : N \rightarrow N$ notation $\Theta(g(n))$ denotes the set of functions

$$\Theta(g(n)) = \left\{ \begin{array}{c} f(n): \text{ there exist positive constants } c_1, c_2 \text{ и } n_0 \\ \text{such that } 0 \le c_1 g(n) \le f(n) \le c_2 g(n) \text{ for all } n \le n_0 \end{array} \right\}$$

Function $f(n)$ belongs to set $\Theta(g(n))$ if there exist positive constants $c_1$ and $c_2$ that allow to keep function $f(n)$ between $c_1 g(n)$ and $c_2 g(n)$ for sufficiently large $n$. Since $\Theta(g(n))$ is a set, one can write $f(n) \in \Theta(g(n))$. One also uses the equivalent notation "$f(n) = \Theta(g(n))$".

For all values of $n$ larger than $n_0$ function $f(n)$ is larger or equal to function $c_1 g(n)$ but is less or equal to $c_2 g(n)$. In other words, for every $n \ge n_0$ function $f(n)$ is equal to $g(n)$ modulo some constant factor. One says that $g(n)$ is *asymptotically exact estimate* of $f(n)$.

The definition of set $\Theta(g(n))$ requires that every element $f(n) \in \Theta(g(n))$ be *asymptotically nonnegative*, i.e., for sufficiently large $n$ function $f(n)$ is nonnegative.

Hence, $g(n)$ should be asymptotically nonnegative, since otherwise set $\Theta(g(n))$ is empty. Hence we consider all functions used in $\Theta$-notation to be asymptotically nonnegative. This assumption holds for other asymptotic notation introduced below.

### $O$-notation

With $\Theta$-notation a function is asymptotically bounded from above and below. If one needs only an *asymptotic upper bound*, one uses $O$-notation. For a given function $g(n)$ notation $O(g(n))$ (pronounced as "*o big of g of n*") denotes the set of functions such that

$$O(g(n)) = \left\{ \begin{array}{c} f(n): \text{ there exist positive constants} c \text{ and } n_0 \\ \text{such that } 0 \le f(n) \le c g(n) \text{ for all } n \le n_0 \end{array} \right\}$$

$O$-notation is used when one needs to give an upper bound of the function modulo some constant factor. For all $n$ larger than $n_0$ one has $f(n) \le c g(n)$. To denote that $f(n)$ belongs to $O(g(n))$ one writes down $f(n) = O(g(n))$. Obviously, $f(n) = \Theta(g(n))$ implies $f(n) = O(g(n))$. In set-theoretic notation $\Theta(g(n)) \subset O(g(n))$. Hence, the fact that $an^2 + bn + c$, where $a > 0$, belongs to set $\Theta(n^2)$ denotes that any square function of this form belongs to set $O(n^2)$. Any linear function $an + b$ for $a > 0$ also belongs to set $O(n^2)$. This can be checked by choosing $c = a+ \mid b \mid$ and $n_0 = max(1, -b/a)$.

**$\Omega$-notation**

*Asymptotical lower bounds* are given in terms of $\Omega$-notation in the same way as $O$-notations formalize asymptotic upper bounds. For a given function $g(n)$ expression $\Omega(g(n))$ (pronounced as "omega big of $g$ of $n$" or simply "omega of $g$ of $n$") denotes set of function such that

$$\Omega(g(n)) = \left\{ \begin{array}{l} f(n): \text{ there exist positive constants } c \text{ and } n_0 \\ \quad \text{such that } 0 \le cg(n) \le f(n) \text{ for all } n \le n_0 \end{array} \right\}$$

For all $n \ge n_0$ one has $f(n) \ge cg(n)$. For any two functions $f(n)$ and $g(n)$ the relation $f(n) = \Theta(g(n))$ holds iff $f(n) = O(g(n))$ and $f(n) = \Omega(g(n))$. Usually this fact is used to define an asymptotical correct estimate by means of asymptotical upper and lower bounds.

$\Omega$-notation is used for expressing lower bounds of time complexity in the best case, hence it gives lower bounds for arbitrary input.

**$o$-notation**

$o$-notation is used to express the fact that upper bound is not asymptotically exact. Formally, set $o(g(n))$ (pronounced as "$o$ small of $g$ of $n$") is defined as follows:

$$o(g(n)) = \left\{ \begin{array}{l} f(n): \text{ for any positive constant } c \text{there exists} \\ n_0 > 0 \text{ such that } 0 \le f(n) < cg(n) \text{ for all } n \ge n_0 \end{array} \right\}$$

Example. $5n^2 = o(n^3)$, but $5n^3 \ne o(n^3)$

Definitions of $O$-notation and $o$-notation are similar. The main difference is that the definition $f(n) = O(g(n))$ bounds function $f(n)$ be the inequality $0 \le f(n) \le cg(n)$ only for some constant $c > 0$, whereas definition $f(n) = o(g(n))$ bounds it by inequality $0 \le f(n) < cg(n)$ for all constants $c > 0$. If $f(n) = o(g(n))$, then $f(n)$ is neglibly small compared to $g(n)$ for $n$ tending to infinity:

$$\lim_{n \to \infty} = \frac{f(x)}{g(x)} = 0.$$

**$\omega$-notation**

By analogy, $\omega$-notation is related to $\Omega$-notation in the same way as $o$-notation is related to $O$-notation. By means of $\omega$-notation one defines a lower bound which is not exact. One can define $\omega$-notation in the following way:

$$f(n) \in \omega(g(n)) \text{iff} g(n) \in o(f(n))$$

Formally $\omega(g(n))$ (pronounced as "omega small of $g$ of $n$") is defined as set

$$\omega(g(n)) = \left\{ \begin{array}{c} f(n): \text{ for any positive constant } c \text{there exists} \\ n_0 > 0 \text{ such that } 0 \le cg(n) < f(n) \text{ for all } n \ge n_0 \end{array} \right\}$$

For example, $\frac{n^2}{2} = \omega(n)$, but $\frac{n^2}{2} \ne \omega(n^2)$. The notation $f(n) = \omega(g(n))$ means that $\lim\limits_{n \to \infty} = \frac{f(x)}{g(x)} = \infty$ if this limit exists. So, $f(n)$ becomes arbitrarily large as compared with $g(n)$ when $n$ tends to infinity.

**Problems, reducibility, complexity classes**

In this section we introduce standard definitions of complexity of algorithms and problems, see [20, 13].

A fundamental notion of the theory of computational complexity is that of *mass problem (generic instance)*, which means a general question for which an answer should be found. Usually a problem has several parameters or free variables with undefined values. Problem $\Pi$ is specified by the following information:

(1) general list of all parameters

(2) formulation of the properties that the *answer* or *solution of the problem* should match.

An instance $I$ is obtained from a mass problem $\Pi$ if all parameters are assigned particular values.

Theory of NP-completeness considers only *decision problems*, which have only two possible answers, "yes" and "no." Formally, a decision problem $\Pi$ consists of two sets, set $D_\Pi$ of all possible *instances* and set of instances $Y_\Pi \subset D_\Pi$ with answer "yes" (yes-instances). The standard form for describing a problem consists of two parts. The first part contains description of the problem and the second part formulates the question which allows for two answers, "yes" and "no." According to tradition the words INSTANCE and QUESTION are typeset in capital letters.

An instance belongs to $D_\Pi$ if it can be obtained from the standard form of a generic instance by substituting constants for all parameters of the condition of the generic instance. An instance belongs to $Y_\Pi$ iff the answer to the question is "yes." Formally, decision problems can be represented by formal languages. For any finite set of symbols $\Sigma$ the set of all finite sequences (words) composed of symbols from $\Sigma$ is denoted by $\Sigma^*$. An arbitrary subset $L \subseteq \Sigma^*$ is called a *language* over alphabet $\Sigma$.

The correspondence between decision problems and languages is established by means of *encoding schemes*. For a given problem $\Pi$, an encoding scheme $e$ describes an instance from $\Pi$ by a word in a fixed alphabet $\Sigma$. So, the problem $\Pi$ and its encoding scheme $e$ partitions set $\Sigma^*$ into three sets: the set of words that are not codes of instances of $\Pi$, the set of words that are codes of instances of $\Pi$ for which the answer is "no", and the set of words that are codes of instances of $\Pi$ for which the answer is "yes." The third set of words constitutes the language $L[\Pi, e]$, which corresponds to problem $\Pi$ under encoding $e$. Every decision problem has an associated encoding-independent function Length: $D_\Pi \to \mathbf{Z}^+$, which gives a formal measure of the *input length (size)* For every "reasonable encoding scheme" $e$, which assumes "natural brevity" or the absence of exponential compressibility, the function Length is *polynomially equivalent* to the length of encoding of an instance, i.e., there exist two polynomials $p$ and $p'$ such that if $I$ is an instance from $D_\Pi$ and word $x$ is the code of $I$ under encoding $e$, then $\text{Length}[I] \leq p(|x|)$ and $|x| \leq p'(Length[I])$, where $|x|$ is the length of word $x$.

The *time complexity* of an algorithm, i.e., a program for deterministic Turing machine or DMT-program, for solving instance $I \in \Pi$ (in particular, $I \in D_\Pi$) with input length (size) $n$ is defined as the number of computation steps before the termination of the program. The *worst-case time complexity* of an algorithm for solution of problem $\Pi$ is defined as maximal time complexity among all instances of size $n$. The worst-case time complexity is denoted by $T_M(n)$. The *worst-case space complexity* of an algorithm for solution of problem $\Pi$ is defined as the maximal amount of memory needed for the algorithm among all instances with input size $n$. Space complexity will be denoted by $S_M(n)$.

Obviously, worst-case complexities give rough estimates of algorithm efficiency. A more adequate estimate could be given by complexity "on average," however in practice theoretical complexity on average is very difficult to compute, since it is usually very hard to obtain an analytical expres-sion for the distribution function for algorithmic complexities of problem instances.

An algorithm (a program for Turing machine) is called polynomial if there exists a polynomial $p$ such that for all $n \in \mathbf{N}$ $T_M(n) \leq p(n)$.

The class of *polynomially computable functions* $P$ is defined as a set of languages

$P = \{L: \text{there exists a polynomial algorithm } M \text{ such that } L = L_M\}.$

**Example 2.1.** Consider a list of objects with unique identifiers from a linearly ordered set $(K; \leq)$, e.g., the set of natural numbers. The *problem of sorting* consists of reordering the list in accordance with the linear order $(K; \leq)$.

*Sorting by choice* starts with finding the object with the least identifier, placing it in the beginning of the list, and iterating these actions until the set is reordered. This algorithm takes $O(n^2)$ time, where $n$ is the size of the list. The advantage of the algorithm is the small amount of data relocation.

Algorithms that have lower polynomial (especially, linear and sublinear) complexity are considered to be scalable for big data applications (although there can be issues related to large constants staying behind $O$ notation) in contrast to exponential algorithms, which need exponentially much time and/or space.

**Example.** Consider set $A = \{a_1, \ldots, a_n\}$ and set $F$ of some subsets of $A$. The problem is to compute all possible intersections of sets from $F$. If $F = \{A \setminus a_i \mid i \in \{1, \ldots n\}\}$, then any subset of $A$ can be obtained as intersection of some sets from $F$. The number of all subsets of $A$ is $2^n$, so in the worst case any algorithm would spend at least $O(2^n)$ time computing and outputting the intersections.

Till now we spoke of complexity of algorithms. Another important issue in algorithmic complexity is that of problem complexity. What is the fastest algorithm for solving a problem? What is the lowest worst-case complexity of an algorithm for solving a particular problem? These questions are often very hard to answer. In particular, there is a famous class of NP-complete problems, for which no polynomial-time algorithms are known, but it is not yet proved mathematically that such algorithms do not exist.

In order to consider these problems, we need to introduce the notion of *nondeterministic Turing machine (NDMT)*, which consists of two modules: usual Turing machine used as *checking module* and a *guessing module*.

Computing with NDMT consists of two stages: the *stage of guessing* and the *stage of checking*:

1) At the guessing stage, upon receiving instance $T$ the guessing module guesses some structure $S \in \Sigma^*$; 2) $I$ and $S$ are input to the checking module, which works like a usual program for Turing machine, it either outputs "yes" (if $S$ is a solution to problem $I$) or "no, " or runs without termination.

A *nondeterministic algorithm (NDA)* is defined as a pair ⟨guess object $S$; check $S$ by deterministic algorithm $A$⟩. NDA "solves" decision problem $\Pi$ if the following two properties hold for all individual problems $I \in D_\Pi$:

1) If $I \in Y_\Pi$, then there exists structure $S$, guessing which for input $I$ results in the fact that the checking stage with input $(I, S)$ terminates with answer "yes."

2) If $I \notin Y_\Pi$, then for input $I$ there is no structure $S$, guessing which would guarantee that the checking stage with input $(I, S)$ terminates with answer "yes".

A nondeterministic algorithm for decision problem $\Pi$ is *polynomial time* if there is a polynomial $p$ such that for every instance $I_\Pi \in Y_\Pi$ there is guess $S$ such that for input $(I, S)$ the checking module outputs "yes" in time $p(\text{Length } |I|)$.

Informally, class NP is the class of all decision problems $\Pi$, which under reasonable encoding can be solved by a nondeterministic algorithm in polynomial time.

**Example.** *GRAPH ISMOMORPHISM*

INSTANCE. Two graphs $G = (V, E)$ and $G' = (V, E')$.

QUESTION. Is it true that $G$ and $G'$ are isomorphic, i.e. there exists bijection $f : V \to V$ such that $\{u, v\} \in E$ iff $\{f(u), f(v)\} \in E'$? $\qquad\square$

*Polynomial transformation (by Carp) of decision problems*

Decision problem $\Pi_1$ is *polynomially transformed* to decision problem $\Pi_2$ (denoted by $\Pi_1 \propto \Pi_2$) if there exists function $f : D_{\Pi_1} \to D_{\Pi_2}$ satisfying the following two conditions:

1. There exists a polynomial DTM program (algorithm) that computes $f$;

2. For all $I \in D_{\Pi_1}$ one has $I \in Y_{\Pi_1}$ iff $f(I) \in Y_{\Pi_2}$.

Decision problem $\Pi_1$ and decision problem $\Pi_2$ are *polynomially equivalent* if $\Pi_1 \propto \Pi_2$ and $\Pi_2 \propto \Pi_1$.

A decision problem is NP-*hard* if any decision problem from NP can be transformed to it. A decision problem is NP-complete if it belongs to NP and is NP-hard. Informally, an NP-complete problem is a hardest (nonunique) problem in class NP. The classical handbook on NP-completeness is the book [20], as well as the "Ongoing guide of NP-complete problems" supported by D.Johnson in Journal of Algorithms. In what follows we will use the following useful properties which hold for arbitrary decision problems $\Pi_1$, $\Pi_2$, $\Pi_3$:

1. If $\Pi \in$ NP, then there exists a polynomial $p$ such that $\Pi$ can be solved by a deterministic algorithm of the time complexity $O(2^{p(n)})$.

2. If $\Pi_1 \propto \Pi_2$ then $\Pi_2 \in P$ implies $\Pi_1 \in P$.

3. If $\Pi_1 \propto \Pi_2$ and $\Pi_2 \propto \Pi_3$, then $\Pi_1 \propto \Pi_3$.

4. If $\Pi_1$ and $\Pi_2$ belong to NP, $\Pi_1$ is NP-complete and $\Pi_1 \propto \Pi_2$, then $\Pi_2$ is NP-complete.

**Example.** Consider some classical NP-complete problems.

1. *3-SATISFIABILITY* (3-SAT)

INSTANCE. 3-CNF, i.e., conjunction of disjunctions $C = \{c_1, c_2, \ldots, c_m\}$ over a finite set $U$ of variables is given so that $|c| = 3, \quad 1 \leq i \leq m$.

QUESTION. Is there a satisfying truth assignment for variables from $U$ so that $C$ become a tautology, i.e., all disjunctions from $C$ become tautologies?

2. *SUBGRAPH ISOMORPHISM*

INSTANCE. Two graphs $G = (V_1, E_1)$ and $H = (V_2, E_2)$.

QUESTION. Does graph $G$ contain a subgraph isomorphic to graph $H$?

In other words, are there subsets $V$, $E$, and function $f : V_2 \rightarrow V$ $V \subseteq V_1, E \subseteq E_1$ such that $|V| = |V_2|, |E| = |E_2|$, and $\{u, v\} \in E_1$ iff $\{f(u), f(v)\} \in E_2$?

3. *CLIQUE*

INSTANCE. A graph $G = (V, E)$ and a natural number $J \leq |G|$.

QUESTION. Is there a subgraph of $G$ that is a clique of size not less than $J$, i.e., a subset $V' \subseteq V$ such that $|V'| \geq J$ and any two vertices from $V'$ are connected by an edge from $E$? $\square$

It is generally believed that the most plausible relation between complexity classes is $P \neq NP$.

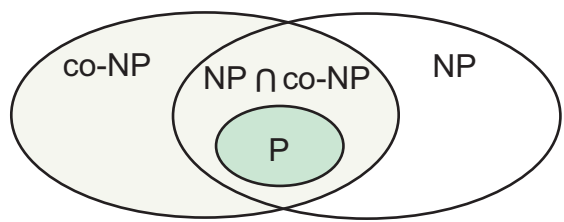

Other possible variants of relations between complexity classes.

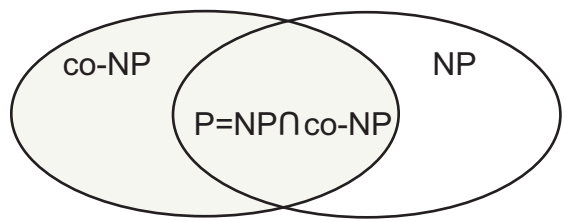

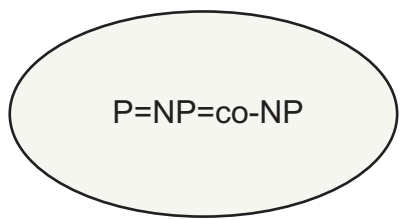

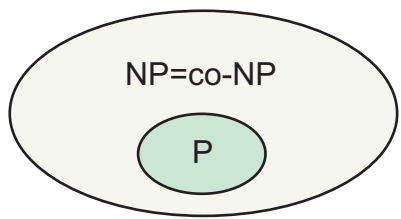

Now from decision problems we come to the enumeration problems where the general question is not "does there exist," but the question "how many exist"? Questions of this kind often arise when allocating computation time and memory for the output, especially when the output can be large, e.g., exponential in the size of the input.

A *search problem* $\Pi$ consists of set $D_\Pi$ of finite objects called instances, and a set (possibly empty) $S_\Pi[I]$ of finite objects, called *solutions*, for every instance $I_\Pi \in D_\Pi$. An algorithm *solves* a search problem $\Pi$ if for an arbitrary instance $I_\Pi \in D_\Pi$ as input it outputs "no" if set $S_\Pi[I]$ is empty or outputs a solution $s \in S_\Pi[I]$.

Note that an arbitrary decision problem can be formulated as a search problem, although intuitively it is not easier to obtain a solution to a problem than to prove its existence. To obtain this reformulation put $S_\Pi[I] = \{\text{"yes"}\}$ if $I_\Pi \in Y_\Pi$, and $S[I] = \emptyset$ otherwise (i.e., if $I_\Pi \notin Y_\Pi$).

An *enumeration problem* based on decision problem $\Pi$ is formulated as follows: "given an instance $I$, what is the size of set $S_\Pi(I)$, i.e., how many solutions does it contain?" Following the work [34] where the class #P was introduced we write down an enumeration problem as follows:

INPUT $I$.

OUTPUT $S_\Pi(I)$.

**Example.**

INPUT a 3-CNF, conjunctive normal form

OUTPUT The number of satisfying assignments for instance 3-SAT.

INPUT Monotone (without negation) CNF with two variables in each disjunction $C = \bigwedge_{i=1}^{s} (x_{i_1} \vee x_{i_2})$, $x_{i_1}, x_{i_2} \in X = \{x_1, \ldots, x_n\}$ for all $i = \overline{1, s}$.

OUTPUT The number of binary vectors of size $n$ (corresponding to assignments of Boolean variables) that satisfy $C$. $\qquad\qquad\square$

Enumeration problem $E$ belongs to class #P if there exists a nondeterministic algorithm such that for every instance of a decision problem $I_\Pi \in D_\Pi$ the number of different "guesses" that lead to "yes" is $|S_\Pi(I)|$, and the duration of the longest check that outputs "yes" is bounded by a polynomial in Length$[I]$.

Class #P can be defined in terms of counting Turing machines as it was done in [34]. Counting Turing Machine (CTM) is an NDTM with additional output band where a special head writes in binary notation the number of accepting computations corresponding to the input. CTM has complexity $f(n)$ in the worst case if the longest accepting computation (for a usual NDTM) for input $n$ is performed in $f(n)$ operations.

#P is a class of functions that can be computed by counting Turing machine with polynomial complexity. Class #P, as well as class NP has complete problems, however transformations in #P are defined slightly different.

A search problem $E_1$ is transformed (by Turing) to search problem $E_2$ (denoted $E_1 \propto_T E_2$) if there exists algorithm $A$ that solves $E_1$ using as a subroutine algorithm $S$ for solution of problem $E_2$, so that if $S$ is a polynomial algorithm for solving $E_2$ then $A$ is a polynomial algorithm for solving $E_1$. Obviously, Carp transformation is a partial case of Turing transformation, where a subroutine is used only once, however it is not known whether the inverse statement holds as well.

Computational problem $E_1$ is #P-*complete* if $E_1 \in$ #P and for all $E_2 \in$#P, $E_2 \propto_T E_1$.

Obviously, the enumeration problems that correspond to NP-complete problems are NP-hard, however there are many decision problems with polynomial algorithms such that the respective enumeration problems are #P-complete, see examples in [35]. It is not yet proved whether #P = P or #P $\neq P$. The problem #P = P? can remain open even if its proved that NP = P, which is highly unlikely. Stating #P-completeness of an enumeration problem implies that the computation of the respective number is fairly hard, in particular, not easier (in terms of Turing transformation) than the problem of the number of satisfying assignments for a CNF.

In what follows we will need definitions of efficiency of algorithms with outputs that can be exponential in the size of their inputs. The following definition was proposed in [25]. An algorithm generating a family of objects called *combinatorial structures* has *delay d* if it satisfies the following conditions for inputs of size $p$:

1. It makes at most $d(p)$ computation steps before outputting the first object or halting without outputting anything.

2. Upon outputting an object it either makes at most $d(p)$ computation steps before outputting the next object or halts.

An algorithm with delay bounded from above by a polynomial in the input size is called *polynomial-delay algorithm* [25].

There are weaker notions of efficiency than polynomial delay. If output of algorithm $A$ with the worst-case time complexity $T_A$ has size $s$ than *amortized time complexity* is $\frac{T_A}{s}$. *Total complexity* of an algorithm is the worst-case time of the algorithm computation expressed as a function of input and output sizes. An algorithm is said to have *polynomial total complexity* if there is a polynomial $p(x, y)$ such that the worst-case time complexity of the algorithm can be bounded from above by $p(i, o)$, where $i$ is the size of input and $o$ is the size of output. Obviously, polynomial delay implies polynomial amortized complexity, and polynomial amortized complexity implies polynomial total complexity.

In [21] a natural generalization for incremental algorithms was proposed. An algorithm that generates a family of objects is called an algorithm with *cumulative delay d* if for any time moment for any input of size $p$ the total number of computer operations did not exceed $d(p)$ plus the product of $d(p)$ with the number of objects output so far. Polynomial cumulative delay means that $d(p)$ is a polynomial of $p$. Total polynomial complexity implies polynomial cumulative complexity.

# Exercises

1. What is the worst-case complexity of computing the inverse relation in case where the relation is represented a) by the set of pairs of the relation b) by the matrix of the relation?

2. What is the worst-case complexity of computing the product of relations in case where the relation is represented a) by the set of pairs of the relation b) by the matrix of the relation?

3. What is the worst-case complexity of computing the transitive closure of a relation in case where the relation is represented a) by the set of pairs of the relation b) by the matrix of the relation?

4. What is the worst-case complexity of checking properties from Definition 1.6 in case where the relation is represented a) by the set of pairs of the relation b) by the matrix of the relation?

5. What is the worst-case complexity of topological sorting? Compare with the complexity of sorting a subset of natural numbers.

6. Show that the transformability by Cook and by Turing define quasi-orders on the set of generic problems.

7. Show that the following problem of computing dimension of a partial order is NP-complete:

   INSTANCE Partially ordered set $(P, \leq)$, natural number $K \in N$.

   QUESTION Is $K$ order dimension of $(P, \leq)$?

8. What is the delay of the algorithm computing elements of a relation on set $A$, $|A| = n$, which is the product of two relations given by binary matrices of size $n \times n$?

# 3 Ordered sets

Order relations play important role in mathematics and data science. Methods of data analysis of different nature are based on the idea of constructing a general description of sets of objects. For example, clustering methods unite objects in clusters that have more general descriptions than objects, because cluster descriptions match more than one object. In supervised learning one computes generalized descriptions of classes based on description of examples. On the one hand, in solving a problem of data analysis, researcher encounters a hierarchical relations of the type "belong to a class," "is a," "be a more general description than," which reflect taxonomies of subject domains. On the other hand, there are important hierarchical relations of the type "to be a part" describing *meronomy* (Greek, $\mu\epsilon\rho o\zeta$, a part) of the subject domain. Both taxonomies and meronomies are naturally described by order relations.

## 3.1   Quasi-order and partial order

**Definition 3.1.** Quasi-order *is reflexive and transitive binary relation.*

**Example 3.1.** Quasi-orders are quite often in applications. Consider the following examples of quasi-order.

1. Entailment relation $\vdash$ on formulas of various logical languages, such as propositional calculus or first-order calculus, is obviously reflexive and transitive, but not antisymmetric. For example,

$$\bar{x} \vee (x \rightarrow y) \vdash \bar{x} \vee y \bar{x} \vee y \vdash \bar{x} \vee (x \rightarrow y).$$

However formulas $\bar{x} \vee (x \rightarrow y)$ и $\bar{x} \vee y$ are (syntactically) different, which can be important, e.g., from the viewpoint of hardware realization.

2. Analysis of consumer baskets. Let $M$ be the set of goods (items) in a supermarket and $C$ be the set of customers during a particular day. Customers can be described by tuples of length $|G|$, where the $i$–th component of the tuple stays for $i$-th item and the value of the component stays for the amount bought by the customer. Customer $x$ bought more or equal than customer $y$ if all components of $x$-tuple are not less than all components of $y$-tuple. The relation "bought more or equal than" on the set of customers is reflexive, transitive, but not antisymmetric, since equality of a purchase does not mean identity of customers.

3. "subgraph isomorphism" relation is obviously reflexive and transtive. If graph $\Gamma_1 = (V_1, E_1)$ is isomorphic to a subgraph of graph $\Gamma_2 = (V_2, E_2)$ and graph $\Gamma_2$ is isomorphic to a subgraph of graph $\Gamma_1$, then graphs $\Gamma_1$ and $\Gamma_2$ are isomorphic, however they do not coincide, since they have different sets of vertices and edges.

In applications one often uses graphs with labeled vertices and edges, such as molecular graphs, semantic networks, parse trees of sentences, XML descriptions, etc.

A *labeled (weighted) graph* is a tuple of the form $\Gamma := ((V, lv), (E, le))$, , where $lv \colon v \to L_v$ и $le \colon e \to L_e$ are labeling functions, which take a vertex and an edge to their labels in sets $L_v$ and $L_e$, respectively.

Labeled graph $\Gamma_1 := ((V_1, lv_1), (E_1, le_1))$ is **isomorphic to a subgraph** of labeled graph $\Gamma_2 := ((V_2, lv_2), (E_2, le_2))$ or $\Gamma_1 \leq \Gamma_2$,

if there exists a bijection $\varphi \colon V_1 \to V_2$ such that

- it defines isomorphism of unlabeled graph $(V_1, E_1)$ to a subgraph of unlabeled graph $(V_2, E_2)$
- respects labels of vertices and edges: $lv_1(v) = lv_2(\varphi(v))$, $le_1(v, w) = le_2(\varphi(v), \varphi(w))$.

Testing subgraph isomorphism of labeled graphs underlies the search by a fragment, indexing and other important operations in databases and knowledge bases where objects are represented by labeled graphs, such as in databases of chemical data, where molecules are given by graphs with labeled vertices and edges.

Note that the union of quasi-order $\trianglelefteq$ and the inverse relation $\trianglerighteq$, which is also a quasiorder, i.e., $\trianglelefteq \cap \trianglerighteq$ is an equivalence relation.

Now we define the following relation on equivalence classes given by quasi-order: for two equivalence classes $\pi$, $\sigma$ one has $\pi \preceq \sigma$ if $p \triangleleft s$ for all $p \in \pi$, $s \in \sigma$. Then the following statement holds.

**Statement 3.1.** Relation $\preceq$ on classes of equivalence given by a quasi-order is reflexive, transitive, and antisymmetric, i.e., it is a partial order.

**Definition 3.2.** Partially ordered set *is a pair* $(P, \leq)$ *where* $\leq$ *is a partial order relation.*

Partially ordered relations are naturally given by acyclic directed graphs. Indeed, if there is a directed cycle in a graph, then due to transitivity and antisymmetry all elements of the cycle should be identical.

**Definition 3.3.** Strict order *is antireflexive, asymmetric, and transitive relation. Strict order* $<$ *related to partial order* $\leq$ *is obtained from* $\leq$ *by deleting pairs of the form* $(a, a)$:

$$x < y := x \leq y \ u \ x \neq y$$

**Example 3.2.** Consider the set of real-valued functions on real numbers: $f \colon R \to R$. For any two functions $g$ and $h$ put $g \leq h$ if $g(x) \leq h(x)$ for all $x \in R$. Thus defined "pointwise" relation on functions $\leq$ is reflexive, transitive, and antisymmetric.

**Example 3.3.** Partial order on partitions.

Recall that *partition* of set $S$ is a set sets (called *blocks*) $\{S_1, \ldots, S_n\}$ such that

$$\bigcup_{i \in [1,n]} S_i = S, \quad \forall i, j \in [1, n] \quad S_i \cap S_j = \emptyset.$$

Usually partitions are denoted as follows: $S_1 \mid S_2 \mid \ldots \mid S_n$.

Partition $P_1$ *is finer* than partition $P_2$ (equivalently, partition $P_2$ *is rougher* than partition $P_1$), or $P_1 \leq P_2$, if for every block $A$ of partition $P_1$ there is block $B$ of partition $P_2$ so that $A \subseteq B$.

**Statement 3.2.** Relation $\leq$ on partitions is a partial order.

For example, for $S = \{a, b, c, d\}$ one has $\{a, b\} \mid \{c\} \mid \{d\} \leq \{a, b, c\} \mid \{d\}$.

**Example 3.4.** Partial order on multisets

**Definition 3.4.** Multiset *on set* $S$ *is set* $S$ *together with function* $r \colon S \to N \cup \{0\}$ *giving multiplicity elements of* $S$.

Many data are naturally represented as multisets, e.g., consumer baskets, molecular formulas, portfolio shares, etc.

Multiset $M$ on $S$ is usually denoted by $\{a_{m_a} \mid a \in M\}$, where $m_a$ is multiplicity of element $a$. Multiset $L = \{a_{l_a} \mid a \in L\}$ is a submultiset of multiset $M = \{a_{m_a} \mid a \in M\}$ ($L \subseteq M$) if for all $a$ one has $l_a \leq m_a$.

For example, for $S = \{a, b, c, d\}$ one has $\{a_1, b_5, c_2\} \subseteq \{a_3, b_6, c_2, d_2\}$.

**Proposition 3.1.** Relation $\subseteq$ on multisets is a partial order.

**Example 3.5.** Partial orders on equivalence classes of logical formulas and classes of isomorphisms of labeled graphs which are obtained from quasi-orders in the way described above.

      The relation "greater or equal" on real numbers.

      The relation "to be a divisor" on natural numbers.

      Containment relation on subsets of a set.

      Real-life examples of partial order are endless. The relation "have to come before" on the set of actions in processes of different types. Collecting documents, cooking scrambled eggs, buying goods in a supermarket. A system of preferences on the set of possible vacation types, and many other relations.

## 3.2  Linear order

A partially ordered set where every two elements are comparable (i.e., $x < y$ or $y < x$) is called *linearly ordered* or *chain*. A partially ordered set where every two elements are incomparable is called an *antichain*.

**Example 3.6.** The sets of natural, rational, real numbers with the standard "greater than or equal" relation is a linear order. A set of classifiers incomparable w.r.t. generality, i.e., having incomparable sets of objects that match the classifiers.

**Example 3.7.** Let $A$ be a finite set of symbols (alphabet) which is linearly ordered by relation $\prec \subseteq A \times A$. A *word* in alphabet $A$ is a finite (may be empty) sequence of symbols from $A$. The set of all words is denoted by $A^*$. *Lexicographic order* $<$ on words from $A^*$ is defined as follows: $w_1 < w_2$ for $w_1, w_2 \in A^*$ if either $w_1$ is a prefix $w_2$ (i.e., $w_2 = w_1 v$, where $v \in A^*$) if the first symbol from the left which is different for $w_1$ and $w_2$ for $w_1$ is less than that of $w_2$ w.r.t. $\prec$. (т.е. $w_2 = wav_1$, $w_1 = wbv_2$, where $w, v_1, v_2 \in A^*$, $a, b \in A$, $a \prec b$).

**Statement 3.3.** Lexicographic order on sets of words $A^*$ is a strict linear order.

Numerous dictionaries in many alphabet-based languages give evidence of this fact.

## 3.3 Order morphisms

**Definition 3.5.** *Mapping $\varphi : M \to N$ between two ordered sets $(M, \leq_1)$ and $(N, \leq_2)$ respects order if for every $x, y \in M$ one has*

$$x \leq_1 y \Rightarrow \varphi x \leq_2 \varphi y.$$

A real-life example of a mapping that respects order is a fair evaluation of material goods in money, where subjective value of goods make usually a partially ordered set for an individual, and amounts of money are linearly ordered like natural numbers.

If the inverse implication holds, i.e. $x \leq_1 y \Leftarrow \varphi x \leq_2 \varphi y$, then $\varphi$ is *order embedding*. Obviously, in the case of monetary evaluation, this condition does not hold (objects having same price are often incomparable w.r.t. their usefulness).

A bijective order embedding is called *order isomorphism*. Not every bijection respecting order is an order isomorphism.

## 3.4 Topological sorting of orders and their representation through linear orders

The ability to enumerate elements of a set, i.e., map its elements to the set of natural numbers helps to set and solve various mathematical problems. A practical example of linear order is money, which allow to map goods and services to linear numbers and compare incomparable things. These properties make linear orders a special class of posets and set problems of relating general partial orders and linear orders. The following theorem of the possibility of *topological sorting* is also known as theorem of Szpilrajn.

**Theorem 3.1.** Let $(S, \leq)$ be a finite poset. Then elements of $S$ can be enumerated in a way such that

$$S = \{s_1, \ldots, s_n\}, \quad s_i \leq s_j \Longrightarrow i \leq j.$$

*Proof.* Let us arbitrarily enumerate the elements of $S$. Take an arbitrary minimal element of $S$ with the least number, let it be element $q$. Put $s_1 := q$. Then consider the poset $(S_1, \leq_1)$, $S_1 := S \setminus \{q\}$, $\leq_1 := \leq \cap S \setminus \{q\}$ and repeat the previous actionswe did for $(S_1, \leq_1)$: find the minimal element $r$ with the least number, put $s_2 := r$, $S_2 := S_1 \setminus \{r\}$, $\leq_2 := \leq_1 \cap S_1 \setminus \{r\}$ and arrive at the poset $(S_2, \leq_2)$. Iterate the procedure until the step $k$ such that $S_k = \emptyset$. The result is the enumeration we are looking for. $\qquad \square$

Both the process and the result of linear ordering of a poset is called *topological sorting*. The result of topological sorting is also called a *linear extension* of the original partial order.

**Theorem 3.2.** Let $(S, \leq)$ be a finite poset where elements $x$ and $y$ are not comparable. Then there are two different topological sortings of $(S, \leq)$ such that $x$ has a greater number than $y$ in one of the sortings and smaller number in the other sorting.

*Proof.* (Idea) The algorithm of topological sorting presented above allows for various orders of enumerating incomparable elements, so both $x$ coming before $y$ and $y$ coming before $x$ are possible. $\qquad \square$

**Example 3.8.**
- There are usually several prerequisites for obtaining an entrance visa to a foregin country: one needs a valid passport, an invitation, and some other documents. Consider the relation "obtaining document $x$ comes before obtaining document $y$". This relation defines a partial order on the starting times of applications for documents. A person who needs to get a visa makes topological sorting of the respective actions w.r.t. linear order on time instances.

- A computation process running on a single processor is based on the precedence relation on intermediate computation results. This relation defines a partial order on time events, which is transformed by the operating system into the linear order of computations executed by the processor.

Obviously, topological sorting of a poset is a mapping that respects order. Topological sorting of a poset (where order is not linear) is an example of bijection, which is not an order isomorphism.

## 3.5 Covering relation and diagram of partial order

**Definition 3.6.** *Interval* $[a, b]$ *is set* $\{x \mid a \leq x \leq b\}$ *of all elements of partial order which lie between $a$ and $b$.*

**Definition 3.7.** *Let* $(P, \leq)$ *be a partially ordered set. Then the respective covering relation $\prec$ is given as follows:*

$$x \prec y := x \leq y, \ x \neq y, \ \nexists z \neq x, y \quad x \leq z \leq y,$$

*or equivalently,*

$$x \prec y := x < y, \ \nexists z \quad x < z < y.$$

**Theorem 3.3.** Let $a < b$ in a poset $(P, \leq)$. Then $P$ contains a subset of elements $\{x_1, \ldots, x_l\}$ such that $a = x_1 \prec \ldots \prec x_l = b$.

*Proof.* by induction on the number of elements $y$ such that $a < y < b$ (i.e., $y$ lies in the interval $[a, b]$). $\qquad\square$

**Theorem 3.4.** Let $(P, \leq_p)$ and $(Q, \leq_q)$ be finite posets with covering relations $\prec_p$ and $\prec_q$, and let $\varphi : P \to Q$ be a bijection. Then the following two statements are equivalent:

1 Bijection $\varphi$ is an order isomorphism, i.e., $x \leq_p y$ iff $\varphi(x) \leq_q \varphi(y)$.

2 $x \prec_p y$ iff $\varphi(x) \prec_q \varphi(y)$.

**Proof.** $1 \to 2$. Let $\varphi$ be an order isomorphism and let $x \prec_p y$, then $x \leq_p y$ and $\varphi(x) \leq_q \varphi(y)$. Suppose that there exists $w$ such that $\varphi(x) <_q w <_q \varphi(y)$. Since $\varphi$ is a bijection, there should exist $u \in P$, a unique preimage of $w$: $\varphi(u) = w$. Since $\varphi$ is an order isomorphism, one has $x <_p u <_p y$, which contradicts the fact that $x \prec_p y$. Hence, $\varphi(x) \prec_q \varphi(y)$.

Now assume that $\varphi(x) \prec_q \varphi(y)$. Since $\varphi$ is an order isomorphism, the inverse mapping $\varphi^{-1}$ will also be an order isomorphism and by reasoning similar to that above we obtain $x \prec_p y$.

$2 \to 1$. Assume that 2 holds and $x <_p y$. By the previous theorem one can find a sequence of covering elements $x = x_0 \prec_p x_1 \prec_p \ldots \prec_p x_n$. By condition 2, $\varphi(x) = \varphi(x_0) \prec_q \varphi(x_1) \prec_q \ldots \prec_q \varphi(x_n)$, hence $\varphi(x) <_q \varphi(y)$. Analogously, due to bijectivity of $\varphi$ and inverse mapping $\varphi^{-1}$, having 2 and $\varphi(x) <_q \varphi(y)$ we obtain $x <_p y$.

**Definition 3.8.** Diagram[1] *of a partially ordered set* $(P, \leq)$ *is a plain geometrical object consisting of circles which centers correspond to elements of the poset and edges that stay for the pairs of the covering relation* $(P, \prec)$, *connect centers of circles and respect the following properties:*

*1.* $a \prec b \Longrightarrow$ *center of the circle corresponding to element a has strictly smaller vertical coordinate than the center of the circle corresponding to element b.*

2. An edge intersects only two circles that correspond to two vertices of the edge.

Remark 1. The possibility of matching condition 1 is guaranteed by the possibility of topological sorting of the poset.

Remark 2. By the previous theorem, partial orders are isomorphic iff they can be represented by same diagram.

Remark 3. By the definition, diagrams cannot contain triangles and horizontal edges.

**Example 3.9.** Graph and diagram of a partial order.

|   | a | b | c | d | e |
|---|---|---|---|---|---|
| a | 1 | 0 | 1 | 1 | 1 |
| b | 0 | 1 | 1 | 1 | 1 |
| c | 0 | 0 | 1 | 0 | 1 |
| d | 0 | 0 | 0 | 1 | 1 |
| e | 0 | 0 | 0 | 0 | 1 |

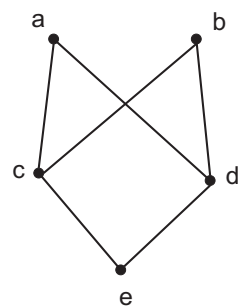

order diagram

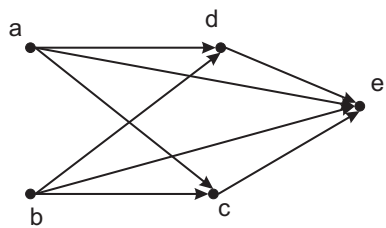

acyclic graph

---

[1]sometimes called Hasse daigram

## 3.6   Duality principle for partial orders

**Definition 3.9.** *Relation $\geq$ inverse to partial order $\leq$ on set $M$, is called* dual partial order $(M, \leq)^d$.

Let $A$ be a statement about partial order $(M, \leq)$. Statement $A^d$ dual to statement $A$ is obtained by replacing $\leq$ by $\geq$.

**Duality principle.** Statement $A$ is valid for poset $(M, \leq)$ if dual statement $A^d$ is valid for dual poset $(M, \leq)^d$.

Duality principle is used for simplification of definitions and proofs: one does not need to prove dual statement if the original statement is already proved. The diagram of the dual poset is obtained from the diagram of the initial poset by "turning it upside down", i.e., using symmetry about the horizontal axis.

## 3.7   Important elements and subsets of posets

**Definition 3.10.** *An element $p \in P$ of partial order $(P, \leq)$ is called* maximal *if there is no element of $P$ strictly larger than $p$: $\forall x \in P \ x \not> p$.*

*Minimal* element is defined dually.

In an arbitrary partial order maximal (minimal) elements are not unique in general.

**Definition 3.11.** *The* largest *element of $(P, \leq)$ is an element $\mathbf{1} \in \mathbf{P}$ that is larger than all other elements: $\forall x \in P \ x \leq \mathbf{1}$.*

The *least* element is defined dually, as the one that is smaller than all other elements.

By definition, if partial order $(P, \leq)$ has a largest (smallest) element, it is unique maximal (minimal) element.

**Definition 3.12.** *Let $(P, \leq)$ be a poset. Subset $J \subseteq P$ is called* (order) ideal *if $x \in J, y \leq x \Rightarrow y \in J$. Dually, subset $F \subseteq P$ is called* (order) filter *if $x \in F, y \geq x \Rightarrow y \in F$.*

In finite case an order filter can be represented by the set of its minimal elements, which are pairwise incomparable, i.e., they form an antichain. Dually, order ideals can be represented by sets of maximal elements, which also make an antichain.

For a subset $Q \subseteq P$ denote $\downarrow Q = \{x \in P \mid (\exists q \in Q) x \le q\}$. Dually, denote $\uparrow Q = \{x \in P \mid (\exists q \in Q) x \ge q\}$. If $Q$ is an antichain, then $\downarrow Q$ is an order ideal, for which $Q$ is the set of maximal elements and $\uparrow Q$ is an order filter, for which $Q$ is the set of minimal elements. If $Q = \{x\}$ for $x \in P$, then the order filter is called *principle filter*.

**Example 3.10.** Order filter and order ideal.

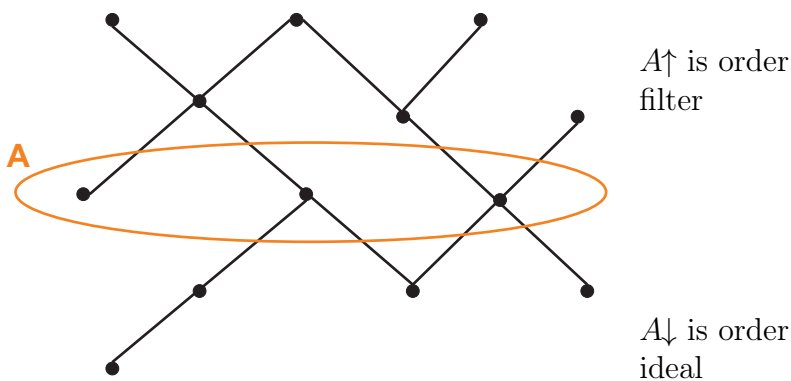

$A\uparrow$ is order filter

$A\downarrow$ is order ideal

The set of order ideals of poset $(P, \le)$ is usually denoted by $O(P)$. This set is partially ordered w.r.t. containment.

**Theorem 3.5.** Let $(P, \le)$ be a poset, then the following three statements are equivalent:

- $x \le y$

- $\downarrow \{x\} \subseteq \downarrow \{y\}$

- $(\forall Q \in O(P)) y \in Q \Leftrightarrow x \in Q$

We let the proof of the theorem to the reader as a simple exercise.

## 3.8  Dimension of a poset

Partial order has "more complicated" structure than a linear order, since one needs several linear orders to represent it adequately, as indicated by the following theorem, called Dushnik-Miller theorem.

**Theorem 3.6.** Any intersection of linear orders on a set is a partial order, and any partial order is an intersection of linear orders that are its linear extensions.

*Proof.* 1. Let $(A, \leq_i)$, $i \in I$ be linear orders on set $A$ and set of indices $I$. Consider their intersection, relation $R$. All linear orders are reflexive: $(a, a) \in (A, \leq_i)$, hence $(a, a) \in R$. If $xRy$ and $yRz$, then $x \leq_i y$, $y \leq_i z$ for any partial order $\leq_i$, therefore, $x \leq_i z$ in all linear orders and $xRz$ in their intersection, i.e., $R$ is transitive. Similar for antisymmetricity: assume that $xRy$ and $yRx$, then for all linear orders one has $x \leq_i y$, $y \leq_i x$, which, by antisymmetricity of linear orders contradicts the fact that $x$ and $y$ are different. Hence, $R$ is antisymmetric.

2. Let $(A, \leq)$ be a partial order, then it is contained (as a set of pairs making relation $\leq$) in all its linear extensions (topological sortings) $(A, \leq_i)$. By 1, the intersection of all $(A, \leq_i)$ is a partial order. This intersection contains partial order $(A, \leq)$, since it is contained in all linear orders $(A, \leq_i)$. It remains to show that the intersection of $(A, \leq_i)$ does not contain elements not belonging to $(A, \leq)$. Assume the converse: there exist two different elements $x$ and $y$ such that the pair $(x, y)$ is not contained in $(A, \leq)$, but it is contained in the intersection of $(A, \leq_i)$, hence, in all linear orders $(A, \leq_i)$. The pair $(y, x)$ cannot be contained in $(A, \leq)$, since by antisymmetry this would imply $x = y$. So, it remains to consider the possibility that $x$ and $y$ are incomparable in $(A, \leq)$. The latter fact implies that there exists a linear extension of $(A, \leq)$ containing $(y, x)$ and not containing $(x, y)$. Hence, the intersection of all linear extensions of $(A, \leq)$ would not contain $(x, y)$, which contradicts our assumption. $\square$

Dushnik-Miller theorem allows one to introduce the following definition of order dimension.

**Definition 3.13.** Order dimension *of partial order* $(P, \leq)$ *is the least number of linear orders* $(P, \leq_i)$, $i \in \{1, \dots, k\}$ *on $P$ so that their intersection gives* $(P, \leq)$: $(P, \leq) = \bigcap_{i \in \{1, \dots, k\}} (P, \leq_i)$.

Another definition of order dimension is based on the product of linear orders.

**Definition 3.14.** Multiplicative dimension *of partial order* $(P, \leq)$ *is the least number $k$ of linear orders* $(P, \leq_i)$, $i \in \{1, \dots, k\}$ *on $P$ such that* $(P, \leq)$ *is order embedded in the Cartesian product* $\times_{i \in \{1, \dots, k\}} (P, \leq_i)$.

**Theorem 3.7.** Order and multiplicative dimensions of a poset coincide.

*Proof.* If partial order $(P, \leq)$ can be represented as intersection of linear orders $\times_i (P, \leq_i)$, $i \in \{1, \ldots, k\}$, then $(P, \leq)$ can be order embedded in the product of these orders, hence the multiplicative dimension is not larger than the order dimension.

In the other direction, let the partial order $(P, \leq)$ be order embeddable in the Cartesian product $\times_i (P, \leq_i)$, $i \in \{1, \ldots, k\}$ of linear orders. We define lexicographically (hence, linearly) ordered set $(P, \leq_i^*)$ for every $i$ in the following way:

$$w \leq_i^* v \iff w_i >_i v_i \text{ or } w_i = v_i \text{ и } w_j > v_j \text{ for the first } j : w_j \neq v_j.$$

On the one hand, $\leq_i^*$ is a linear order. On the other hand, due to embedding of the initial poset $(P, \leq)$ in the product of linearly ordered sets $\times_i (P, \leq_i)$ the condition $w_j > v_j$ for the first $j : w_j \neq v_j$ denotes that $w_j \geq v_j$ for all other $j$.

It can be easily tested that the initial poset $(P, \leq)$ can be embedded in the intersection of linear orders $(P, \leq_i^*)$, hence order dimension of $(P, \leq)$ is not larger than its multiplicative dimension. $\square$

Dushnik-Miller theorem and the definition of the order dimension can be naturally applied to the analysis of preferences: for every finite poset of alternatives there exists a set of linearly ordered sets ("scales") where all alternatives are comparable and the intersection of the scales gives the original order. The scales can be considered as different aspects of alternatives, w.r.t. which all alternatives are pairwise comparable. For example, in selecting a car, these scales can be price, year of production, producer, model, color, etc.

**Example 3.11.** Determining order dimension of a poset.

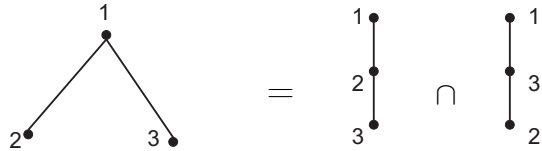

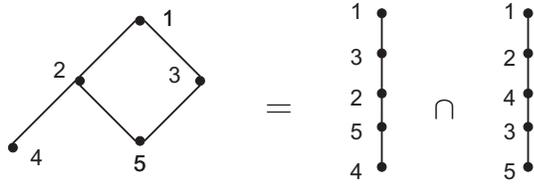

**Example 3.12.** Determining multiplicative dimension of a poset.

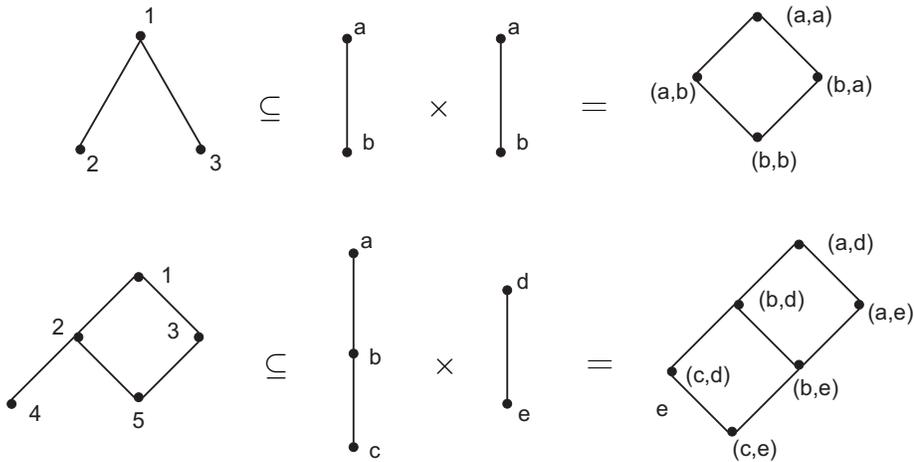

Dushnik-Miller theorem gives a representation of a partial order through intersections of linear orders. An alternative view – through the union of orders – is given by the following theorem of Dilworth.

The *width* of a partial order is the size of its maximal antichain.

**Theorem 3.8.** Let $(P, \leq)$ be a poset of width $k$. Then set $P$ can be represented as a partition $P = P_1 \cup \ldots \cup P_k$, $i \neq j \Leftrightarrow P_i \cap P_j = \emptyset$, where each block $P_i$ is linearly ordered w.r.t. $\leq$.

The proof of the theorem of Dilworth requires elaboration of the theory of matchings, so we refer the reader to special literature on the subject, e.g. [32].

# Exercises

1. Show that the incomparability relation for a poset is a tolerance.

2. Show that subgraph isomorphism relation is a quasi-order on the set of graphs.

3. Show that the entailment relation on first-order predicate logic is a quasi-order.

4. Show that Carp and Turing reductions between computation problems are quasi-order relations on the set of computation problems.

5. Show that any subset of a poset is a poset.

6. Is a strict order antisymmetric?

7. Show that $(P^c)^d = (P^d)^c$ for any partiall ordered relation $P$.

8. Show that the relation inverse to a partial order is a partial order.

9. Show that any finite nonempty poset has a minimal element.

10. Show that the intersection of partial orders on a set is a partial order.

11. Compute diagram of a partial order given by relation matrix.

12. Show that the diagram of a partial order cannot contain triangles and horizontal lines.

13. Show that lexicographic order is a strict linear order.

14. Construct diagrams of all partial orders on a set of 4 (5) elements.

15. Let $A$ be a nonempty set and $P$ be the set of all partial orders on $A$. For $\rho, \sigma \in P$ define $\rho \leq \sigma$ if $a\rho b$ implies $a\sigma b$. Show that $(P, \leq)$ is a poset.

16. Construct the diagram of a partial order on partitions of a 4-element set.

17. Show that if $(P, \leq)$ is an antichain, then $O(P)$, the set of order ideals of $P$, is the powerset of $P$.

# 4 Closures and lattices

## 4.1 Semilattices and lattices as partially ordered sets

**Definition 4.1.** *Let $(P, \leq)$ be a poset and $A \subseteq P$.* Upper bound *of a set $A \subseteq P$ is set*

$$\{b \in P \mid \forall a(a \in A \rightarrow b \geq a)\}.$$

**Definition 4.2.** Supremum (least upper bound) *of set $A \subseteq P$, denoted* sup $(A)$, *is the least element $b$ of $A$ (if it exists):*

1. *$\forall a(a \in A \rightarrow b \geq a)$,*

2. *$\forall x \forall a((x \in P) \& (a \in A \rightarrow x \geq a) \rightarrow x \geq b)$.*

The notion of *infimum (greatest lower bound*, denoted by *inf($A$)*, is introduced dually. In the literature on order theory supremums and infimums are also called *joins* and *meets*, respectively.

*Join and meet of the emptyset.* Condition 1 of the definition of supremum allows an arbitrary element to be a supremum of an emptyset, since $a \in \emptyset$ is false for every $a \in P$ and the implication of condition 1 is always true. In condition 2 the internal implication $\forall a \forall x(a \in \emptyset \rightarrow x \geq a)$ is a tautology and for the external implication to be true, it is necessary that $x \geq b$ for every $x$, which is possible only when $b$ is the least element of the partial order. Hence, $\sup(\emptyset)$ exists if there is a least element in $(P, \leq)$. Similarly, $\inf(\emptyset)$ is the largest element of the partial order if it exists.

**Definition 4.3.** *A partially ordered set $(SL, \leq)$ is an* upper semilattice *if any pair of elements $x, y \in SL$ has supremum $\sup\{x, y\}$.*

The definition of *lower semilattice* is dual.

**Definition 4.4.** *A partially ordered set $(SL, \leq)$ is a* lower semilattice *if any pair of elements $x, y \in SL$ has infimum $\inf\{x, y\}$.*

**Example 4.1.** Examples of semilattices given by diagrams:

Lower semilattice                    Upper semilattice

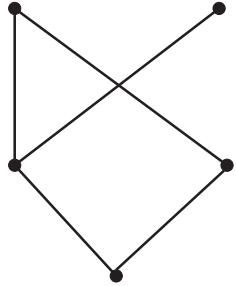
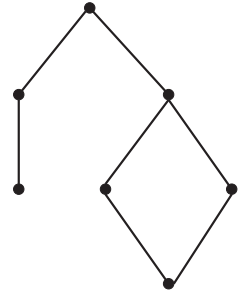

**Definition 4.5.** *A partially ordered set* $(L, \leq)$ *is a* lattice *if any pair of elemenets* $x, y \in L$ *has supremum* $\sup\{x, y\}$ *and infimum* $\inf\{x, y\}$.

**Example 4.2.**

A poset which is neither a lower, nor    A poset, which is a
an upper semilattice                      lattice

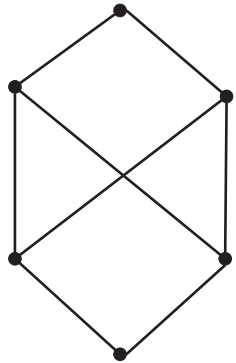
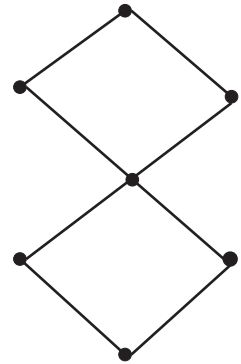

Many standard posets introduced above are lattices.

The set of natural numbers, partially ordered by "less than or equal" relation, is a lattice: the join of two numbers is their maximum and the meet is their minimum.

The relation of divisibility on natural numbers defines a lattice, where each pair of natural numbers has join, which is their least common multiple, and meet, which is their greatest common divisor.

The set of all subsets of a set, ordered by containment relation, is a lattice, where the meet is the intersection and the join is the union of two sets.

The set of partitions of set $S$ ordered by the reflexive relation "to be a rougher partition than" makes a lattice, called partition lattice. The join of two partitions $A$ and $B$ consists of the blocks that are unions of all blocks from $A$ and $B$ that are not disjoint (i.e., having non-empty intersection). The meet of $A$ and $B$ consists of inclusion-maximal blocks, each of which is a non-empty intersection of blocks from $A$ and $B$.

Lattices of order filters and indeals. For an arbitrary poset $(P, \leq)$ the set of order ideals $\mathrm{OI}(P, \leq)$ is partially ordered by set-theoretic containment. This order makes a lattice. Indeed, let $I_1$ and $I_2$ be two arbitrary ideals from $\mathrm{OI}(P, \leq)$. Then $\inf\{I_1, I_2\}$ is the set-theoretic intersection $I_1 \cap I_2$, and $\sup\{I_1, I_2\}$ is the set-theoretic union $I_1 \cup I_2$.

**Example 4.3.** Partition diagram of a four-element set.
$$A = \{1, 2, 3, 4\}$$

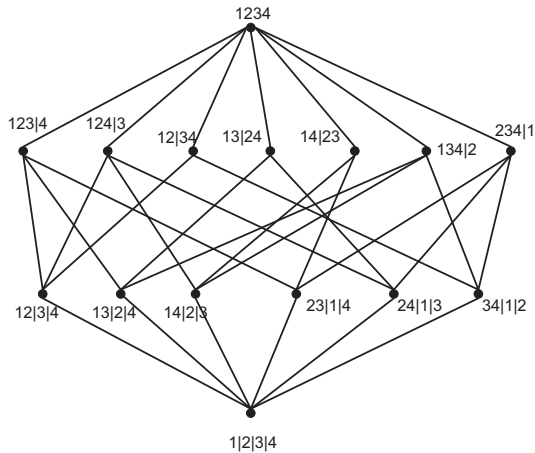

## 4.2 Lattices as algebras

**Theorem 4.1.** An arbitrary set $L$ is a lattice w.r.t. a partial order iff operations $\vee$ and $\wedge$ with the following properties are defined for any $x, y, z \in L$:

L1    $x \vee x = x, \quad x \wedge x = x$   (idempotence)

L2    $x \vee y = y \vee x, \quad x \wedge y = y \wedge x$   (commutativity)

L3    $x \vee (y \vee z) = (x \vee y) \vee z, \quad x \wedge (y \wedge z) = (x \wedge y) \wedge z$   (associativity)

L4    $x = x \wedge (x \vee y) = x \vee (x \wedge y)$   (absorption)

*Proof.* Necessity. Let $L$ be a lattice w.r.t. a partial order $\leq$. We take sup as $\wedge$ and inf as $\vee$, and show that $\wedge$ and $\vee$ satisfy properties L1-L4. Indeed, properties L1, L2 hold due to the fact that an element can occur in a set only once and the order of elements in the set is inessential (commutativity). Then, by the definition of supremum

$$\sup\{x, \sup\{y, z\}\} \geq x, y, z,$$

hence

$$\sup\{x, \sup\{y, z\}\} \geq \sup\{x, y\}, \sup\{x, \sup\{y, z\}\} \geq z$$

and

$$\sup\{x, \sup\{y, z\}\} \geq \sup\{\sup\{x, y\}, z\}.$$

Similarly, in the other direction

$$\sup\{\sup\{x, y\}, z\} \geq \sup\{x, \sup\{y, z\}\}.$$

Hence,

$$\sup\{\sup\{x, y\}, z\} = \sup\{x, \sup\{y, z\}\},$$

and property L3 holds. By the definition of infimum, $x \geq \inf\{x, y\}$, hence, by the definition of supremum,

$$\sup\{x, \inf\{x, y\}\} = x,$$

i.e., property L4 also holds.

Sufficiency. Let operations $\vee$, $\wedge$ with properties L1-L4 be given on $L$. We define the following relation $\leq$ on elements of $L$: $x \leq y := x \wedge y = x$. Note that $x \leq y$ iff $x \vee y = y$. Indeed, let $x \leq y$, then $x \wedge y = x$ and $x \vee y = (x \wedge y) \vee y = y$ by property L4. Inversely, if $x \vee y = y$, then $x \wedge y = x \wedge (x \vee y) = x$ by property L4 and $x \leq y$ holds.

We show that $L$ is a lattice w.r.t. $\leq$, where $\sup\{x, y\}$ and $\inf\{x, y\}$ are defined as $x \vee y$ and $x \wedge y$, respectively. Indeed, first $x \vee y \geq y$, since $x \vee y \wedge y = y$. Second, let $z \geq x$ and $z \geq y$. Then $z \vee x = z$, $z \vee y = y$ и $z \vee (x \vee y) = (z \vee x) \vee y = z \vee y = z$, hence $z \geq (x \vee y)$ and $x \vee y$ is the join of $x$ and $y$. Similarly, we show that $x \wedge y$ is the meet of elements $x$ and $y$ w.r.t. partial order $\leq$. $\square$

This theorem allows one to consider a lattice as an algebra $(L, \vee, \wedge)$ with properties L1-L4.

**Definition 4.6.** Natural order *of a lattice given as algebra with properties L1-L4 is the relation* $\leq \subseteq L \times L$ *defined as* $x \leq y \overset{def}{=} x \wedge y = x$ *(or, equivalently, as* $x \leq y \overset{def}{=} x \vee y = y$).

## 4.3 Complete lattices

**Definition 4.7.** *A lower semilattice is called* complete *if its any subset $X$ (e.g., the empty set) has infimum* $\bigwedge X$.

Complete upper semilattice is defined dually . It is easily seen that an upper semilattice has the largest element (the unit of the upper semilattice) equal to the supremum of the set of all lattice elements, and the lower semilattie has the least element (lattice zero) equal to the infimum of all elements of the semilattice.

**Definition 4.8.** *A lattice is called* complete *if every subset of it, including the empty one, has supremum and infimum.*

Note that a complete semilattice is a complete lattice. Indeed, consider a complete semilattice w.r.t. operation $\bigwedge$, then we define operation $\bigvee$ as follows: $\bigvee X := \bigwedge\{z \mid z \geq x \text{ for all } x \in X\}$. By definition, $\bigwedge\{z \mid z \geq x, z \geq y\} \geq x$ for any $x \in X$ and for any $z \geq x$ such that $x \in X$ by the definition of infimum $\bigwedge X \geq z$, hence $\bigvee X$ is the supremum of $X$.

By the definition, even the empty subset of elements of a complete lattice has infimum. As we showed in Section 4.1, supremum of the empty set is the smallest element (zero) of the lattice. Dually, infimum of the empty set is the largest element (unit) of the lattice.

$$\bigvee \emptyset = \mathbf{0} \qquad \bigwedge \emptyset = \mathbf{1}$$

Hence, a complete lattice has zero and unit, i.e., every complete lattice is bounded.

**Statement 4.1.** Obviously, all finite lattices are complete. Indeed, supremum and infimum of the emptyset are equal to zero and unit of the lattice, resepectively, while supremum and infimum of any nonempty subset is defined by pairwise supremums and infimums: for $X = \{x_1, \ldots, x_n\}$ one can set $\bigwedge X = x_1 \wedge x_2 \wedge \ldots \wedge x_n$ independent of the order of elements and brackets, by associativity and commutativity of operation $\wedge$.

We present an important example of a complete lattice, which will be used below.

**Definition 4.9.** *If $X$ is a set and $L \subseteq 2^X$ such that $X \in L$ and for any nonempty subset $A \subseteq L$ one has $\bigcap \{a \in A\} \in L$, then $L$ is called* closure system (Moore family) *over $X$.*

**Statement 4.2.** A closure system $L$ over $X$ is a complete lattice, where infimum and supremum are given in the following way:

$$\bigwedge_{i \in I} A_i = \bigcap_{i \in I} A_i,$$

$$\bigvee_{i \in I} A_i = \bigcap \{B \in L \mid \bigcup_{i \in I} A_i \subseteq B\}.$$

In Data Science an important example of a closure system is a set of descriptions of some objects from a domain together with all possible intersections of such descriptions. These intersections give similarity of objects in terms of common parts of descriptions. Such closure systems will be studied in detail in the section on Formal Concept Analysis.

Closure systems are tightly related to closure operators.

**Definition 4.10.** Closure operator *on set $G$ is a mapping $\varphi\colon \mathcal{P}(G) \to \mathcal{P}(G)$ that takes every subset $X \subseteq G$ to its closure $\varphi X \subseteq G$ and has the following properties:*

1. *$\varphi\varphi X = \varphi X$ (idempotence)*

2. *$X \subseteq \varphi X$ (extensivity)*

3. *$X \subseteq Y \Rightarrow \varphi X \subseteq \varphi Y$ ( monotonicity )*

Elements $X \subseteq G$ is called *closed* if $\varphi X = X$.

The following statement connects closure operators and closure systems.

Let a set $X$ be given, then 1. a set of subsets of $X$ closed w.r.t. some closure operator $\varphi(X)$ makes a closure system over $X$. 2. 2. A closure system defines a closure operator where all elements of the closure system are closed w.r.t. the closure operator.

Proof. 1. Let $A$ and $B$ be closed sets, i.e., $\varphi(A) = A$, $\varphi(B) = B$.

Then, by the monotonicity of closure operator, $\varphi(A \cap B) \subseteq \varphi(A) = A$, $\varphi(A \cap B) \subseteq \varphi(B) = B$, hence, $\varphi(A \cap B) \subseteq A \cap B$. By the extensivity of a closure operator, $A \cap B \subseteq \varphi(A \cap B)$, hence $\varphi(A \cap B) = A \cap B$, i.e., the intersection of closed sets is a closed set, q.e.d.

2. Let $L$ be a closure system over set $X$. Define operator $\varphi(A) = \cap_{A \subseteq X, X \in L} X$. Obviously, $\varphi(A)$ is monotonic and extensive. We show that it is also idempotent. Consider $\varphi\varphi(A) = \cap_{\varphi(A) \subseteq X, X \in L} X$. On the one hand, due to the extensivity, one has $A \subseteq \varphi(A)$, hence, $\varphi(A) \subseteq X$ implies $A \subseteq X$. On the other hand, by the definition of $\varphi(A)$, if $A \subseteq X$, then $\varphi(A) \subseteq X$. Hence, we have the equivalence of the conditions $A \subseteq X \varphi(A) \subseteq X$ and $\varphi\varphi(A) = \varphi(A)$, i.e., the idempotence is proved. Thus, $\varphi(A)$ is a closure operator such that all sets from the closure system $L$ are closed with respect to it.

In computer science there are numerous examples of closure operators. An example from data science can be taken from data mining, where concise representation of association rules is based on so-called closed itemsets (closed sets of attributes). Suppose that we have a dataset given by sets of objects described by itemsets (sets of attributes). Support of an itemset $B$ is the set of objects with descriptions containing $B$. $B$ is a *closed itemset* if $B \cup \{a\}$ has smaller support than $B$ for any $a \notin B$. A mapping taking an itemset $A$ to the smallest closed itemset $B$ containing $A$ is a closure operator.

## 4.4 Duality principle for lattices

A statement dual to a statement on lattices can be obtained by replacing symbols $\leq$, $\vee$, $\wedge$, 0, 1 with symbols $\geq$, $\wedge$, $\vee$, 1, 0, respectively.

If poset $(V, \leq)$ is a (complete) lattice, then a dual partially ordered set $(V, \leq)^d = (V, \geq)$ is also a (complete) lattice called dual to the initial one. A diagram of a dual lattice can be obtained from a diagram of the initial lattice by "turning it upside down."

## 4.5 Sublattices

**Definition 4.11.** *A triple $\mathcal{K} = (K, \wedge, \vee)$ is a* sublattice *of lattice $\mathcal{L} = (L, \wedge, \vee)$ if $K$ is a nonempty subset of set $L$ with the following property: $a, b \in K$ implies that $a \wedge b \in K$, $a \vee b \in K$, where $\wedge$ and $\vee$ are taken in $L$, i.e., they are restrictions of operations of $\mathcal{L}$ to $K$.*

**Example 4.4.** The lattice of binary vectors of dimension 3 and its sublattice. Binary vectors of a fixed dimension make a lattice w.r.t. componentwise disjunctions and conjunctions.

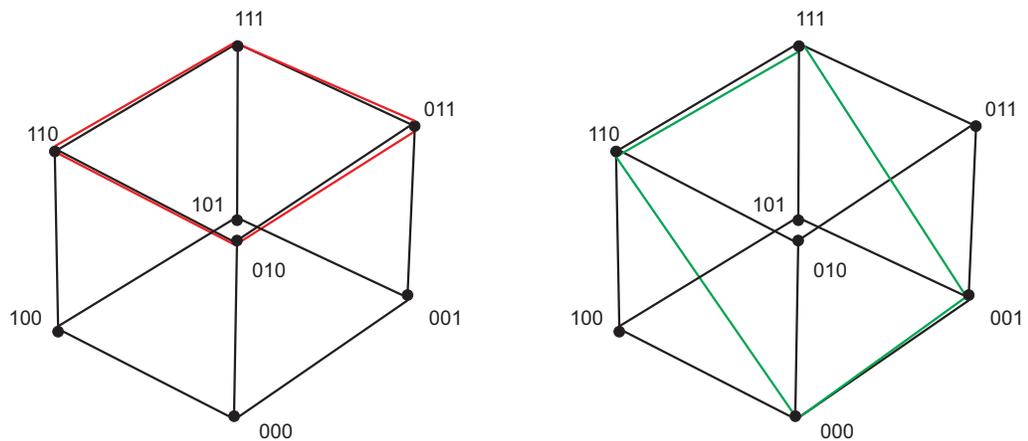

## 4.6   Lattice morphisms

**Definition 4.12.** *A mapping between two complete lattices* respects supremums *(is a* join-morphism *) if*

$$\varphi \bigvee X = \bigvee \varphi(X)$$

$\wedge$-morphism is defined dually.

**Definition 4.13.** *A* complete homomorphism *(* homomorphism of complete lattices*) is a mapping between two complete lattices that is supremum and infimum morphism.*

*Isomorphism of complete lattices* is a bijective complete homomorphism.

## 4.7   Distributivity and modularrity

**Definition 4.14.** *A lattice whith the following properties*

$$x \wedge (y \vee z) = (x \wedge y) \vee (x \wedge z)$$

$$x \vee (y \wedge z) = (x \vee y) \wedge (x \vee z)$$

*is called* distributive.

**Example 4.5.** *Ring of sets* over a set $I$ is a set $F$ of subsets of $I$ that, together with two sets $S$ and $T$ contains their set-theoretic intersection and union $S \cup T$.

**Definition 4.15.** *A lattice with the property*

$$if \ x \leq z, \ then \ x \vee (y \wedge z) = (x \vee y) \wedge z$$

*is called* modular.

**Definition 4.16.** Pentagon *is a lattice with the following diagram:*

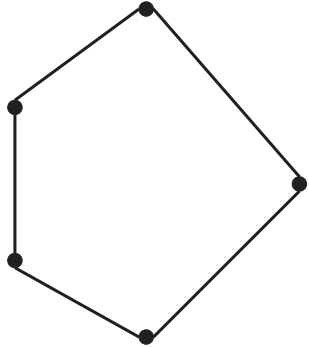

**Definition 4.17.** Diamond *is a lattice with the following diagram:*

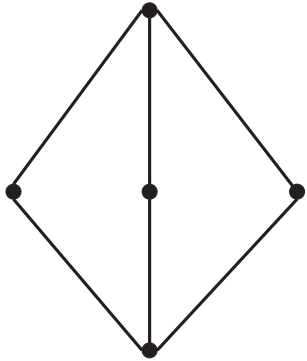

Lattice $L$ contains pentagon (diamond) if it contains a sublattice which is a pentagon (diamond). The following theorem was proved by G. Birkhoff [?].

**Theorem 4.2.**  1. A lattice is distributive iff it contains neither a pentagon, nor a diamond.

  2. A lattice is modular iff it does not contain a pentagon.

  The proof of necessity of theses statements is easy, we let it as an exercise to the reader. The proof of sufficiency requires introduction of new definitions and is beyond the scope of this book. The proofs can be found, e.g., in [?].

  A Boolean lattice is a very important type of lattices, where, besides supremums and infimums, one has operation called complement.

**Definition 4.18.** *Let $L$ be a bounded lattice, i.e., it has zero $0$ and unit $1$. For $a \in L$ elements $\bar{a} \in L$ is called* complement *if $a \wedge \bar{a} = 0$ and $a \vee \bar{a} = 1$.*

**Definition 4.19.** *Lattice $L$ is called Boolean if it is*

1. *distributive,*

2. *bounded, i.e., if it has zero $0$ and $1$,*

3. *every element $a \in L$ has a unique complement $\bar{a} \in L$.*

The proof of the following statement, which characterizes Boolean algebras as algebras with three operations is left to the reader as an exercise.

**Statement 4.3.** If $L$ is a Boolean lattice, then the following relations hold:

1. $\bar{0} = 1$, $\bar{1} = 0$

2. $\bar{\bar{a}} = a$

3. $\overline{(a \vee b)} = \bar{a} \wedge \bar{b}$, $\overline{(a \wedge b)} = \bar{a} \vee \bar{b}$

4. $a \wedge \bar{b} = 0$ iff $a \leq b$ $a, b \in L$.

## 4.8 Irreducible elements of a lattice

Irreducible elements play similar important role for lattices as prime numbers play for natural numbers.

Let $L$ be a lattice. Element $x \in L$ is called $\vee$-*irreducible* (pronounced supremum- (or join-) irreducible) if

1. $x \neq 0$ (in case where $L$ has zero),

2. If $x = a \vee b$, then $x = a$ or $x = b$.

$\wedge$-*irreducible* (infimum- or meet-irreducible) elements are defined dually. Irreducible elements do not exist necessarily in infinite lattices.

In diagrams of finite lattices irreducible elements are easy to see: a $\vee$-irreducible element has only one neighbor from below and a $\wedge$-irreducible element has only one neighbor from above.

A join (meet) reducible element is an element that is not join (meet-) irreducible, so it can be represented as a join (meet) of all elements below (above) it.

# Excercises

1. Show that in a finite lattice the idempotence of both operations (L1) follows from the other properties (L2−L4).

2. Show that distributivity implies absorption.

3. Show that for any three elements $x$, $y$, $z$ of a lattice one has inequality $x \vee (y \wedge z) \leq (x \vee y) \wedge (x \vee z)$.

4. Show that for any three elements $x$, $y$, $z$ of a lattice one has inequality $x \wedge (y \vee z) \geq (x \wedge y) \vee (x \wedge z)$.

5. Show that for any three elements $x$, $y$, $z$ of a lattice the equalities $x \wedge (y \vee z) = (x \wedge y) \vee (x \wedge z)$ and $x \vee (y \wedge z) = (x \vee y) \wedge (x \vee z)$ are logically equivalent.

6. Show that the conditions of idempotence, extensivity, and monotonicity for a closure operator are equivalent to a single condition $y \cup \overline{y} \cup \overline{\overline{x}} \subset \overline{x \cup y}$, where $\overline{(\cdot)}$ is a closure operator.

7. Show that for any lattice the operations of infimum and supremums are isotonic, i.e., $y \leq z \rightarrow x \wedge y \leq x \wedge z$ и $x \vee y \leq x \vee z$.

8. Show that for any four elements $x$, $y$, $z$, and $q$ of a lattice the inequality $(x \vee y) \wedge (z \vee q) \geq (x \wedge z) \vee (y \wedge q)$ holds.

9. Show that if in a distributive lattice one has $c \wedge x = c \wedge y$ и $c \vee x = c \vee y$, then $x = y$.

10. Show that pentagon and diamond are not distributive lattices.

11. Show that

    a) the set of all subintervals of interval [0,1] b) the set of all subtrees of a tree

    are closure systems.

# 5 Introduction to Formal Concept Analysis

## 5.1 Main definitions

**Definition 5.1.** *Let two sets $G$ and $M$ be given, they are considered as sets of objects and attributes, respectively. Let $I \subseteq G \times M$ be a binary relation between $G$ and $M$. $gIm$ or $(g, m) \in I \iff$ object $g$ has attribute $m$. Triple $\mathbb{K} := (G, M, I)$ is called a* (formal) *context.*

A context is naturally represented by a binary table, where units stay for elements of relation $I$, or by a bipartite graph, where vertices of different parts stay for objects and attributes, and edges stay for the elements of $I$.

Let a context $\mathbb{K} := (G, M, I)$ be given, consider two mappings

$$\varphi \colon 2^G \to 2^M \; \varphi(A) \stackrel{\text{def}}{=} \{m \in M \mid gIm \;\text{ for all } g \in A\}.$$

$$\psi \colon 2^M \to 2^G \quad \psi(B) \stackrel{\text{def}}{=} \{g \in G \mid gIm \;\text{ for all } m \in B\}.$$

For any $A_1, A_2 \subseteq G$, $B_1, B_2 \subseteq M$

1. $A_1 \subseteq A_2 \Rightarrow \varphi(A_2) \subseteq \varphi(A_1)$

2. $B_1 \subseteq B_2 \Rightarrow \psi(B_2) \subseteq \psi(B_1)$

3. $A_1 \subseteq \psi\varphi(A_1)$ и $B_1 \subseteq \varphi\psi(B_1)$

**Definition 5.2.** *It is easily seen that a pair of mappings $\varphi$ and $\psi$ that satisfy conditions above defines a* Galois connection *between $(2^G, \subseteq)$ and $(2^M, \subseteq)$, i.e., $\varphi(A) \subseteq B \Leftrightarrow \psi(B) \subseteq A$.*

According to FCA traditions, instead $\varphi$ and $\psi$ one uses notation $(\cdot)'$ (so-called Darmstadt notation), so that $(\cdot)''$ denotes both the composition $\varphi \bigcirc \psi$ and the composition $\psi \bigcirc \varphi$.

For arbitrary $A \subseteq G$ and $B \subseteq M$

$$A' \stackrel{\text{def}}{=} \{m \in M \mid gIm \text{ for all } g \in A\},$$
$$B' \stackrel{\text{def}}{=} \{g \in G \mid gIm \text{ for all } m \in B\}.$$

**Definition 5.3.** *A* (formal) *concept is a pair $(A, B)$: $A \subseteq G$, $B \subseteq M$, $A' = B$, $B' = A$.*

Set $A$ is called *(formal) extent*, and $B$ is called *(formal) intent* of concept $(A, B)$. Concepts are partially ordered by generality relation $(A_1, B_1) \geq (A_2, B_2) \iff A_1 \supseteq A_2 \quad (B_2 \supseteq B_1)$, which is read as "concept $(A_1, B_1)$ is more general than concept $(A_2, B_2)$."

A subset of attributes of the form $\{g\}', g \in G$ is called *object intent* and a subset of objects of the form $\{m\}', m \in M$ is called *attribute extent*.

In the binary table of the context a concept corresponds to an inclusion-maximal (not necessarily maximum) subtable filled with ones. In the bipartite graph $(G \cup M, I)$ of the context a concept corresponds to an inclusion-maximal complete bipartite subgraphs (maximal bicliques), also not necessarily maximum.

**Example 5.1.** Example of a context

| G \ M | a | b | c | d |
|-------|---|---|---|---|
| 1. △ | × | | | × |
| 2. ◺ | × | | × | |
| 3. ▭ | | × | × | |
| 4. □ | | × | × | × |

Objects:

**1** equilateral triangle,

**2** right triangle,

**3** rectangle,

**4** square,

Attributes:

**a** has exactly 3 vertices,

**b** has exactly 4 vertices,

**c** has a right angle,

**d** equilateral.

## 5.2 Properties of operations $(\cdot)'$

Let $(G, M, I)$ be a formal context, $A, A_1, A_2 \subseteq G$ be subsets of objects, $B \subseteq M$ be a subset of attributes, then

1. If $A_1 \subseteq A_2$, then $A_2' \subseteq A_1'$;

2. If $A_1 \subseteq A_2$, then $A_1'' \subseteq A_2''$;

3. $A \subseteq A''$;

4. $A''' = A'$, hence $A'''' = A''$;

5. $(A_1 \cup A_2)' = A_1' \cap A_2'$;

6. $A \subseteq B' \Leftrightarrow B \subseteq A\prime \Leftrightarrow A \times B \subseteq I$.

Similar relations hold for subsets of attributes.

A simple, but very important consequence of property (5), is the fact that the set of all common attributes of objects from set $A \subseteq G$ is the intersection of object intents of all objects from $A$: $A\prime = \cap_{g \in A}\{g\}$.

Recall the definition of the closure operator and closure system from the previous chapter:

**Example 5.2.** Let context $(G, M, I)$ be given, then operators $(\cdot)''\colon 2^G \to 2^G$ and $(\cdot)''\colon 2^M \to 2^M$ are closure operators, the set of all intents and the set of all extent make closure systems.

## 5.3   Basic Theorem of Formal Concept Analysis

The basic theorem of Formal Concept Analysis consists of two parts: The first part says that the ordered set of formal concepts of a context is a lattice, this fact is based on the proof of the respective properties of polarities by G. Birkhoff [?]. The second part gives properties of a lattice that allow it to be isomorphic to a concept lattice. For finite lattices the second part was proved in [6]. The modern formulation of the basic theorem of FCA was given in [38], see also [19].

To formulate the theorem we need the following definition.

**Definition 5.4.** *A subset $X \subseteq L$ for a lattice $(L, \leq)$ is called* supremum-dense *if any element of the lattice $v \in L$ can be represented as*

$$v = \bigvee \{x \in X \mid x \leq v\}.$$

Similarly we introduce the notion of *infimum-dense* subset.

**Theorem 5.1.** The partial order of concepts of context $(G, M, I)$ is a complete lattice [19, 37]. It is called the concept lattice of context $(G, M, I)$ and denoted by $\underline{\mathfrak{B}}(G, M, I)$. For an arbitrary subset of formal concepts

$$\{(A_j, B_j) \mid j \in J\} \subseteq \underline{\mathfrak{B}}(G, M, I)$$

Infimums and supremums are given as

$$\bigwedge_{j \in J} (A_j, B_j) = \left( \bigcap_{j \in J} A_j, \left( \bigcup_{j \in J} B_j \right)'' \right), \tag{1}$$

$$\bigvee_{j \in J} (A_j, B_j) = \left( \left( \bigcup_{j \in J} A_j \right)'', \bigcap_{j \in J} B_j \right). \tag{2}$$

Complete lattice $V$ is isomorphic to the lattice $\underline{\mathfrak{B}}(G, M, I)$ iff there exist mappings $\gamma \colon G \to V$ and $\mu \colon M \to V$ such that $\gamma(G)$ is supremum-dense in $V$, and $\mu(M)$ is infimum-dense in $V$, and $gIm \Leftrightarrow \gamma g \leq \mu m$ for all $g \in G$ and for all $m \in M$. In particular, $V \cong \underline{\mathfrak{B}}(V, V, \leq)$.

*Proof.* 1. We show that formal concepts make a lattice w.r.t. partial order $\leq$ introduced above. Let $(A_1, B_1)$ and $(A_2, B_2)$ be formal concepts of the context $\underline{\mathfrak{B}}(G, M, I)$. Let $(A_1, B_1) \wedge (A_2, B_2) = (A_1 \cap A_2, (A_1 \cap A_2)')$. We show that $(A_1 \cap A_2, (A_1 \cap A_2)')$ is a formal concepts. To this end, it remains to show that $(A_1 \cap A_2)'' = (A_1 \cap A_2))$. Indeed,

$$(A_1 \cap A_2)'' = (B_1' \cap B_2')') = (B_1 \cup B_2)''') = (B_1 \cup B_2)' = (B_1' \cap B_2') = (A_1 \cap A_2).$$

The fact that $\wedge$ is supremum w.r.t. order $\leq$ on concepts follows from the definition of this order and the fact that $\cap$ is supremum w.r.t. order $\subseteq$. Similarly, the statement is proved for the infimum of an arbitrary nonempty set of concepts. The infimum of the empty set of concepts is concept $(G, G')$, which is the largest element of the ordered set of concepts. The statement on the supremum is proved similarly.

2. Idea of the proof for the finite case. Let a finite lattice $V$ be given. We need to create individual labels of lattice elements by moving bottom-up from the smallest element of the lattice, using as elementary labels elements of a set $J$, the labels will be then sets of elements of $J$. The additional requirement of the labeling is that it should be parsimonius, i.e., using as small set of labels from $J$ as possible. Then the bottom element (zero of the lattice) gets the empty set for its label, atoms of the lattice (i.e., upper neighbors of the zero) get one-element labels, and an element lying above atoms will get the union of the labels of its lower neighbors. For elements that have more than one neighbor, due to the uniqueness of supremum in the complete lattice, the labels of elements will also be unique. For elements of the lattice that have only one lower element, we need to take an additional element from $J$ to make its label be different from that of its unique lower neighbor. In such a way we can label all elements of the lattice using the smallest possible set of elements of $J$, so cardinality of $J$ can not exceed the cardinality of $V$.

Similarly, we can label elements of lattice $V$ by moving top-down and using some set $M$ of elementary labels. Again, for this labeling, the cardinality of $M$ cannot exceed the cardinality of $V$.

When both labelings are performed, every element of lattice has a pair of labels $(A, B)$, where $A \subseteq J$, $B \subseteq M$. This pair corresponds to a formal concept of context $(J, M, \leq)$, where relation $\leq$ is given as follows: $j \leq m$, if for the bottom-up labeling the lattice element, where $j$ appeared first is less or equal than the lattice element, where $m$ appeared first for the top-down labeling. □

## 5.4 Duality principle for concept lattices

Denote the sets of supremum- and infimum-irreducible elements of lattice $(L, \leq)$ by $J(L)$ and $M(L)$, respectively. It is easily seen that $J(L)$ and all its supersets are supremum-dense, whereas set $M(L)$ and all its supersets are infimum-dense in $(L, \leq)$.

The basic theorem of FCA implies that

$$L \cong \underline{\mathfrak{B}}(J(L), M(L), \leq).$$

If $(G, M, I)$ is a formal context, then $(M, G, I^{-1})$ is also a formal context and $\underline{\mathfrak{B}}(M, G, I^{-1}) \cong \underline{\mathfrak{B}}(G, M, I)^d$. Moreover, the mapping $(B, A) \to (A, B)$ defines an isomorphism: by interchanging objects and attributes we obtain the dual concept lattice.

## 5.5 Clarification and Reduction of Context

In contexts there can be sets of attributes with same attribute extents and sets of objects with same object intents. Obviously, one can leave only one representative for such a set to obtain an isomorphic concept lattice, but some extents and intents will contain less objects and attributes, respectively. This procedure is called *clarification*. Formally, clarification is factorization of the sets of objects and attributes w.r.t. the equivalence relation $g' = h'$ $(m' = n')$, $g, h \in G$, $m, n \in M$.

Assume also that some attributes are possessed by all objects of the context, i.e., $m' = G$ for all such $m$. Then $m$ belongs to all intents of the context, including the top concept $(G, G')$. If we delete all attributes of this kind, then all intents will change equally (the intent of the concept $(G, G')$ will become empty) and the partial order on concepts, hence supremums and infimums will not change.

Similar for the case where attribute extent $m'$ is equal to the intersection of several attribute extents for some set of attributes $B$: $m$ occurs in a concept intent iff all attributes from $B$ occur. Deleting $m$ results in a lattice isomorphic to the original one, this fact can be used for reducing the complexity of computation. Formally,

**Definition 5.5.** *Attribute $m \in M$, $\mathbb{K} = (G, M, I)$ is reducible if its extent is equal to the intersection of attribute extents for some set of attributes:*

$$m' = G \ or \ m' = \bigcap \{n' \mid n \in M \ \& \ n' \supset m'\}.$$

If $m$ is reducible, then $\underline{\mathfrak{B}}(G, M, I) \cong \underline{\mathfrak{B}}(G, M \setminus \{m\}, I \cap (G \times (M \setminus \{m\})))$.

Reducible objects are defined dually. An irreducible attribute corresponds to the infimum-irreducible element of the concept lattice that is the object concept for this object. Dually, an irreducible object corresponds to supremum-irreducible element of the concept lattice that is the object concept for this object.

**Example 5.3.** In this table attribute $m_k$ is reducible, since $m'_k = m'_i \cap m'_j$

| $G \setminus M$ | $\ldots$ | $m_i$ | $\ldots$ | $m_j$ | $\ldots$ | $m_k$ | $\ldots$ |
|---|---|---|---|---|---|---|---|
| $g_1$ | $\ldots$ | $\times$ | $\ldots$ | | $\ldots$ | | $\ldots$ |
| $g_2$ | $\ldots$ | $\times$ | $\ldots$ | $\times$ | $\ldots$ | $\times$ | $\ldots$ |
| $g_3$ | $\ldots$ | $\times$ | $\ldots$ | $\times$ | $\ldots$ | $\times$ | $\ldots$ |
| $g_4$ | $\ldots$ | | $\ldots$ | $\times$ | $\ldots$ | | $\ldots$ |

## 5.6 Attribute implications

**Definition 5.6.** *For context $K = (G, M, I)$ and sets $A, B \subseteq M$ the expression $A \to B$ is a valid implication if $A' \subseteq B'$, i.e., every object that has all attributes from $A$ also has all attributes from $B$.*

Valid implications can be seen in the diagram of the concept lattice as follows. If $A \rightarrow B$, then the infimum of all attribute concepts of all attributes from $A$ in the diagram of the concept lattice lies below the infimum of all attribute concepts of attributes from $B$.

A subset of attributes $T \subseteq M$ respects (is a model of implication) $A \rightarrow B$ if $A \nsubseteq T$ or $B \subseteq T$. This fact is denoted by $T \models A \rightarrow B$.

A subset of attributes $T \subseteq M$ respects (is a model of) set of implications $\mathcal{L}$ if it satisfies every implication from set $\mathcal{L}$, this is denoted by $T \models \mathcal{L}$.

Implication $A \rightarrow B$ is valid on set $\mathcal{T}$ of subsets of the attribute set $M$ if every subset $T_i \in \mathcal{T}$ respects implication $A \rightarrow B$.

Implication $A \rightarrow B$ (semantically) follows from set of implications $\mathcal{L}$ if every subset of attributes that is a model of $\mathcal{L}$ is also a model of $A \rightarrow B$.

Set of implications $\mathcal{N}$ semantically follows from set of implications $\mathcal{L}$ if every implication from $\mathcal{N}$ semantically follows from $\mathcal{L}$.

The definition of a valid implication implies that all object intents are its models. Moreover, all context intents are models of a valid implication. Indeed, every intent of context $D \subseteq M$ is an intersection of some object intents $D = g'_1 \cap \ldots \cap g'_n$. Since $g'_i$ is a model of $A \rightarrow B$, i.e., $A \nsubseteq g'_i$ or $B \subseteq g'_i$, then $A \nsubseteq D$ or $B \subseteq D$ holds also for the intersection of object intents $D = g'_1 \cap \ldots \cap g'_n$.

Below, if it does not lead to confusion, we write just "implication" instead of "valid implication".

It is easily tested that implications satisfy the following *Armstrong rules*:

$$\frac{}{X \rightarrow X} \quad (1)$$

$$\frac{X \rightarrow Y}{X \cup Z \rightarrow Y} \quad (2)$$

$$\frac{X \rightarrow Y, Y \cup Z \rightarrow W}{X \cup Z \rightarrow W} \quad (3)$$

Indeed, the first Armstrong rule holds because $X' \subseteq X'$. For the second rule: if $X \rightarrow Y$, then $X' \subseteq Y'$, hence $X' \cap Z' \subseteq Y'$ for every $Z$ and $X \cup Z \rightarrow Y$. For the third rule: if $X' \subseteq Y'$, $(Y \cup Z)' = Y' \cap Z' \subseteq W'$, then $(X \cup Z)' = X' \cap Z' \subseteq W'$. Hence, deduction with Armstrong rules is correct, i.e., applying Armstrong rules to valid implications one obtains valid implications.

The following useful implication properties follow from Armstrong rules (1)-(3).

$$\frac{X \subseteq Y}{Y \to X} \quad (4)$$

$$\frac{X_1 \to Y_1, \quad X_2 \to Y_2}{X_1 \cup X_2 \to Y_1 \cup Y_2} \quad (5)$$

$$\frac{X \to Y}{X \to Y \setminus X}, \quad \frac{X \to Y \setminus X}{X \to Y} \quad (6)$$

We leave the proof of the validity of these properties as an exercise for the reader.

**Theorem 5.2.** (Completeness of deduction with Armstrong rules)

If an implication can be deduced by Armstrong rules from set of valid implications $\mathcal{L}$, then it follows semantically from $\mathcal{L}$, since implications obey Armstrong rules.

*Proof.* We show that an implication that follows semantically from a set of implications $\mathcal{L}$ of a context is deducible from set of implications $\mathcal{L}$ by Armstrong rules. Let $\mathcal{L} \models V$ and $\mathcal{L} \vdash Y$, where $Y$ is the maximal subset of attributes such that $X \vdash Y$, in particular, $X \subseteq Y$. Assume that $V \nsubseteq Y$, then $Y$ does not respect $X \to V$, hence, it does not respect $\mathcal{L}$. Therefore, $\mathcal{L}$ contains implication $Z \to W$, which is not respected by $Y$, i.e., $Z \subseteq Y$, but $W \nsubseteq Y$.

Since $Z \subseteq Y$, then $X \to Z$ can be deduced from $X \to Y$ by Armstrong rules. By rule (2), $X \to Z$ and $Z \to W$ imply $X \to W$, which contradicts the maximality of $Y$ and the fact that $W \nsubseteq Y$. Hence, $V \subseteq Y$ and everything that semantically follows from $\mathcal{L}$, is (syntactically) deducible from $\mathcal{L}$. $\qquad \square$

Denote by $\mathcal{L}$ the set of all valid implications of the context. Let $Imp$ be a set of implications, then by $Imp(X)$ we denote the set of attributes obtained from $X$ in the following way:

$$Imp(X) := X \cup \{B \mid A \subseteq X, A \to B \in \mathcal{L}\}.$$

Denote $X_1 := X, X_n := Imp(X_{n-1})$ and $X^{Imp} := X_k$ for $k$ such that $X_k = Imp(X_k)$. Obviously, operator $(\cdot)^{Imp}$ is closure operator on the set of attributes, which will be called implicational closure w.r.t. set of implications $Imp$, and we say that $Imp(X)$ is *derived* from $X$.

**Theorem 5.3.** The implicational closure coincides with $(\cdot)''$ closure based on Galois connection, i.e., $\mathcal{L}(X) = X''$.

*Proof.* On the one hand, since $X \to X''$ is a valid implication of the context $(X' \subseteq X''' = X')$, we have $X'' \subseteq (X)^{\mathcal{L}}$. On the other hand, it can be easily seen that $X_{n-1} \subseteq X''$ implies $X_n \subseteq X''$ for $Imp = \mathcal{L}\}$. Hence, by induction we obtain $(X)^{\mathcal{L}} \subseteq X''$ and $X'' = (X)^{\mathcal{L}}$. $\qquad\square$

## 5.7 Implication bases

Using Armstrong rules the set of implications can be represented concisely by a subset of implications, from which all other implications can be deduced using Armstrong rules. Here we consider some important complete (but not always irredundant) subsets of $\mathcal{L}$, i.e., one can deduce all implications of the context from them.

### 5.7.1 Generator implicational basis and direct basis

By definition, $A \to B$ implies that $B \subseteq B'' \subseteq A''$, hence, everything that can be derived from subset of attribute $A$ is contained in the closure $A''$, and it suffices to have subset of implications $\{A \to A'' \mid A \subseteq M\}$ for deducing all other implications by means of Armstrong rules. Next, having implication $A \to A''$, we can decrease the premise without losing the conclusion $A''$ till the minimal subset $C \subseteq A$ such that $C'' = A''$.

**Definition 5.7.** *Subset of attributes $D \subseteq M$ is a* generator *of a closed set of attributes $B \subseteq M$, $B'' = B$ if $D \subseteq B$, $D'' = B = B''$.*

Subset $D \subseteq M$ is *minimal generator* if for every $E \subset D$ one has $E'' \neq D'' = B''$, i.e., every proper subset of a minimal generator is a generator of a smaller closed set.

Generator $D \subseteq M$ is called *nontrivial* if $D \neq D'' = B''$. The set of all nontrivial minimal generators of $B$ will be denoted by nmingen$(B)$.

*Generator basis* of implications is defined as follows:

$$\{F \to (F'' \setminus F) \mid F \subseteq M, F \in \text{ nmingen } (F'')\}.$$

Consider the context from Example 5.1 to see that a generator cover can be redundant. Compute minimal generators for all closed sets of attributes (intents). Nontrivial minimal generators are sets $b, bd, cd, ab, acd$. The set of respective implications is $b \to bc, bd \to bcd, cd \to bcd, ab \to abcd, acd \to abcd$. Obviously, implication $bd \to bcd$ is deducible from implication $b \to bc$ and implication $acd \to abcd$ is deducible from implication $cd \to bcd$ by means of the Armstrong rules, which demonstrates the redundancy of the minimal generator basis.

In [**?**, 19] a subset of generators that makes a smaller basis which was proposed.

**Definition 5.8.** *A subset $A \subseteq M$ is a* proper premise *if*

$$A'' \neq A \cup \bigcup_{n \in A} (A \setminus \{n\})''.$$

So, a proper premise is a subset of attributes $A$ such that its closure cannot be deduced from the union of closures of proper subsets of $A$, i.e., $A$ "contains some essential information" that is not contained neither in its proper subsets nor in their union.

Every proper premise is a minimal generator. Indeed, if a generator is not minimal, then one can take its subset to obtain the same closure. However, the inverse does not hold in general. Consider subset $A = bd$ in the above example. It is a minimal generator, but it is not a proper premise, since $A'' = bcd = b'' \cup d'' = A \cup \bigcup_{n \in A} (A \setminus \{n\})''$.

**Theorem 5.4.** The set of implications

$$\mathcal{D} = A \to A'' \mid A \text{ is a proper premise}$$

is complete, i.e., all implications of the context can be deduced from it by means of Armstrong rules.

*Proof.* 1. We show by induction that every implication $A \to A''$ can be deduced from D by means of Armstrong rules. Since for any valid implication $A \to B$ we have $B \subseteq A''$, this will be sufficient to prove the theorem. Consider implications of the form $m \to m''$. If $m$ is not a premise, then one has $\emptyset \to m''$, the empty set is a proper premise and implication $m \to m''$ is obtained from it by Armstrong rule (2). Suppose we have proved the theorem for premises up to size $n - 1$. Consider an implication $A \to A_J$ where $|A| = n$. If $A$ is a proper premise, then the implication is in D. If $A$ is not a proper premise. Consider all $n$ proper subsets of size $n - 1$, $A_1, \ldots A_n$. Then by definition of proper premise we have $A\mathtt{J} = A_{\mathtt{J}1} \cup \ldots \cup A_n''$. However, by the inductive assumption, all implications $A_i \to A_i''$ are deduced from D. Then, by Armstrong rules (see properties above), we have $\bigcup_{i=1,n} A_i \to \bigcup_{i=1,n} A_{\mathtt{J}i} = A_{\mathtt{J}}$, so we have deduced $A \to A''$ from D, hence this set is a cover of implications. □

A *direct implicational basis* is a subset of implications such that all other implications of the context can be obtained "in one deductive step", i.e., $X^{\mathcal{D}} = \mathcal{D}(X)$ for any $X \subseteq M$.

Implicational cover with proper premises is a direct basis, since all other implications of the context can be obtained "in one step": $X^{\mathcal{D}} = \mathcal{D}(X)$ for any $X \subseteq M$, see [8, 1, 36]. Moreover, D is cardinality minimal direct basis, as shown in [8].

**Example 5.4.** For the context from Example 5.1 proper premises are $b, ab, cd$ with respective implications $b \to bc, ab \to abcd, cd \to bcd$. None of the implication is redundant.

Proper basis was reinvented several times, see [8, 1, 36]. For example, in paper [36] the following definition was proposed.

**Definition 5.9.** *Let $B \subseteq M, m \in M$. We call $B$ a stem for $m$, and $m$ a root for $B$, if $B$ is minimal with the property that $m \in B\mathtt{J}$. Further $B \subseteq M$ is a stem if it is a stem for some $m$, and $m \in M$ is a root if it is a root for some $B$. If $B$ is a stem, we put $roots(B) := \{m \in E \mid m \text{ is a root for } B\}$. Similarly, if $m$ is root, we put $stems(m) := \{U \subseteq M \mid B \text{ is a stem for } m\}$. Note that $m \in M$ is not a root iff $M \setminus \{m\}$ is closed. Vice versa, a subset $S$ does not contain a stem iff all subsets of $S$ (including $S$ itself) are closed.*

*Then $\{X \to roots(X) \mid X \subseteq M \text{ is a stem } \}$ is* called canonical direct implicational basis.

The proof of the fact that proper premises are exactly premises of the canonical direct implicational basis is left to the reader as an exercise.

### 5.7.2 Minimal implication base

Cardinality minimal (minimum) implication base (not necessarily a direct one) was proposed in paper [23] in terms of "nodes of irredundancy". An equivalent definition in terms of pseudo-intents was given later, see [19].

A subset of attributes $P \subseteq M$ is called a *pseudo-intent* if $P \neq P''$ and for any pseudo-intent $Q$ such that $Q \subset P$ one has $Q'' \subset P$.

Then the set

$$\mathrm{DG} = \{P \rightarrow (P'' \setminus P) \mid P \text{ is a pseudo-intent }\}$$

is minimum (cardinality minimal) implicational base, called *canonical base, stembase, or Duquenne-Guigues basis.*

Denote by $\mathrm{DG}(X)$ the implicative closure of subset of attributes $X$ w.r.t. set of implications DG.

**Theorem 5.5.** (Completeness of Duquenne-Guigues implication base). Every valid implication of the context can be inferred from the Duquenne-Guigues implication base by means of Armstrong rules.

*Proof.* Assume that DG set of implications is not complete. Then there exists a valid implication $X \rightarrow Y$, which cannot be deduced from DG. In particular, $Y \subseteq X'' \subseteq DG(X)''$, but $Y \not\subseteq DG(X)$. Hence, $DG(X) \neq DG(X)''$ and $DG(X)$ is not closed. Let $Q \subset DG(X)$ be a pseudo-intent which is proper subset of $DG(X)$. Since $Q \rightarrow Q''$, then one should have $Q'' \subseteq DG(X)$. Hence, $DG(X)$ is not closed, but it contains the closure of any pseudo-closed proper subset. Hence, by definition, $DG(X)$ is a pseudo-intent. $\qquad \square$

One can construct pseudo-intents by simple search through the powerset of the set of attributes $M$. A more efficient algorithm will be considered in the next section. For our example pseudo-intents are sets $b, ab, cd$, so in this case the proper premise direct base and Duquenne-Guigues base coincide. If we extend the context by a new attribute $e$ with extent $e' = \{1, 2, 4\}$, then $be$ will be a proper premise, but not a pseudo-intent, since for pseudo-intent $\{b\} \subseteq \{b, e\}$ one has $\{b\}'' = \{b, c\} \not\subseteq \{b, e\}$. This example shows that a direct basis can be larger than Duquenne-Guigues base, but in practice the application of the direct base can be more efficient than the application of Duquenne-Guigues base, since everything deducible from the direct base can be obtained in one pass over the set of implications.

## 5.8 Attribute Exploration

The interactive procedure of attribute exploration is a sequence of alternating steps where odd steps are made by a computer which outputs the implication basis and the even steps are made by a human expert who either accepts the set of output implications as a (partial) theory of the subject domain, or disagrees with some implications by giving new objects that do not respect them. For detailed description of various versions of Attribute Exploration see [18].

## 5.9  Implications and functional dependencies

A *many-valued context* is a quadruple $(G, M, W, I)$, where $G$ is a set of objects, $M$ is a set of attributes, $W$ is a set of attribute values, and $I \subseteq G \times M \times W$ is such that $(g, m, w) \in I$, and $(g, m, v) \in I$ implies $w = v$. Attribute $m$ is *complete* if for all $g \in G$ there exists $w \in W$ such that $(g, m, w) \in I$. A many-valued context is complete if all its attributes are complete. For complete many-valued contexts the value of attribute $m$ for object $g$ is denoted by $m(g)$, so $(g, m, m(g)) \in I$. *Functional dependency* $X \to Y$ holds in complete many-valued context $(G, M, W, I)$ if for every pair of objects $g, h \in G$ one has

$$(\forall m \in X \quad m(g) = m(h)) \Rightarrow (\forall n \in Y \quad n(g) = n(h)).$$

For every multi-valued context $K = (G, M, W, I)$ one can construct context $K_N := (\mathcal{P}_2(G), M, I_N)$, where $\mathcal{P}_2(G)$ is the set of pairs of different objects from $G$, and $I_N$ is defined as

$$\{g, h\} I_N m :\Leftrightarrow m(g) = m(h).$$

Functional dependency $X \Leftrightarrow Y$ holds in a many-valued context $K$ iff implication $X \to Y$ is valid in context $K_N$. The inverse transformation also holds.

**Theorem 5.6.** For context $K = (G, M, I)$ one can construct a many-valued context $K_W$ such that implication $X \to Y$ is valid in $K$ if functional dependency $X \Leftrightarrow Y$ holds in $K_W$.

**Example 5.5.** Consider the following formal context $K = (G, M, I)$:

| = | a | b | c | d |
|---|---|---|---|---|
| 1 | × | × |   |   |
| 2 | × |   | × |   |
| 3 |   | × |   | × |
| 4 | × |   | × |   |

The many-valued context where functional dependencies of $K_W$ have the same syntactical form as implications of $K$ looks as follows:

| = | a | b | c | d |
|---|---|---|---|---|
| 0 | × | × | × | × |
| 1 | × | × | 1 | 1 |
| 2 | × | 2 | × | 2 |
| 3 | 3 | × | 3 | × |
| 4 | × | 4 | × | 4 |

# Exercises

1. For the context of Example 5.1 construct the set of all formal concepts and draw a diagram of the concept lattice.

2. Prove the properties of operators $(\cdot)'$ and $(\cdot)''$.

3. Show that in the concept lattice the set of all object concepts is supremum-dense (recall that for object $g$ the respective object concept is $(\{g\}'', \{g\}')$. A hint: show that every concept is the supremum of all object concepts lying below.

4. Prove the first part of the basic theorem of FCA.

5. Prove that for any formal context $(G, M, I)$ concept lattices $(\underline{\mathfrak{B}}(G, M, I))^d$ and $\underline{\mathfrak{B}}(G, M, I^{-1})$ are isomorphic, i.e., $I^{-1} := \{(m, g) \in M \times G \mid (g, m) \in I\}$.

6. For the context from Example 5.1 construct all implications and the canonical Duquenne-Guisgues basis.

7. Prove the validity of Armstrong rules (1)-(3) and properties (4)-(6) of implications using prime operators $(\cdot)'$.

8. Deduce properties (4)-(6) using Armstrong rules (1)-(3).

9. For example 5.5 show that upon the deletion of row 0 from the many-valued context, the statement of theorem 5.6 would not hold.

10. Prove that the set of proper premises and the set of premises of direct basis coincide, so the respective definitions are equivalent.

# 6 Main algorithms of FCA

## 6.1 Preprocessing algorithms

Before computing concepts, concept lattices, implications, and other useful tool of knowledge discovery that we will consider in the next section, one should prepare data. Preprocessing can consist in scaling, clarification, reduction, and sorting of objects and/or attributes w.r.t. sizes of object intents and/or attribute extents. These procedures are quite simple.

*Scaling of a many-valued context* $(G, M, W, I)$. The algorithmic complexity of scaling differs much depending on the type of scaling. For example, in case of nominal scaling, a binary context is generated with the number of attributes $\sum_{m \in M} |W_m|$, where $W_m$ is the number of all possible values of the many-valued attribute $m$. Filling the entries of the new table takes $O(|G| \cdot \sum_{m \in M} |W_m|)$ time. In case of interordinal scaling a context with $2 \cdot \sum_{m \in M} |W_m|$ attributes is created. When scaling graph descriptions (see section ?? on pattern structures) new binary attributes stay for subgraphs of original graph data, so in the worst case the number of new attributes can be exponential in the size of data.

*Clarification of context* consists in deleting repeated rows and columns from the binary cross-table. Formally, this means that for every equivalence class of objects and attributes w.r.t. equivalence relations $g' = h'$ and $m' = n'$, respectively, one retains a single representative. For object clarification this can be attained by sorting rows (of size $|M|$) in time $O(|M| \cdot |G| \cdot \log |G|)$. Then one needs to pass through all rows, say from the first to the last one, while comparing each row with the preceding one and deleting the current row if it coincides with the preceding one, in time $O(|M| \cdot |G|)$. Hence, object clarification takes $O(|M| \cdot |G| \cdot \log |G|)$ time. Similarly for attributes, where clarification takes $O(|G| \cdot |M| \cdot \log |M|)$ time.

(*Attribute reduction of context*), i.e., deleting reducible attributes of context $\mathbb{K} = (G, M, I)$ can be run as follows: for every attribute column $m$ one looks for every column $n$ containing it, i.e., with $m' \leq n'$. Then, the intersection of these columns is compared with $m$: if the intersection coincides with $m$, then $m$ is reducible, if not, then the attribute is not reducible. Thus, testing reducibility of an attribute takes $O(|M| \cdot |G|)$ time and the entire attribute reduction, i.e., testing reducibility for all attributes takes $O(|M|^2 \cdot |G|)$ time. Dually, object reduction takes $O(|G|^2 \cdot |M|)$ time.

*Sorting by cardinality.* Some algorithms perform much faster when rows and/or columns are sorted by the number of units in them (or cardinality of respective object intents and/or attribute extents). One can count all object intents in time $O(|G| \cdot |M|)$ and then sort object intents w.r.t. these counts in time $O(|G| \cdot \log |G|)$. Sorting w.r.t. attributes takes $O(|G| \cdot |M|)$ and $O(|M| \cdot \log |M|)$, respectively.

## 6.2 Algorithms for generating the set of concepts and covering relation

Many algorithms were proposed for computing formal concepts and the covering relation of the concept lattice (or Galois lattice, treillis de Galois in francophone literature), starting from 1960s. The state of the art by 2002 was presented in [28], where most efficient and popular algorithms were analyzed, like NextClosure [17], algorithm of Bordat [12], Close-by-One (CbO) [27], and some other. After 2002 several new efficient algorithms for computing concepts and the covering relation appeared, the most efficient of them are AddIntent [26], FastCbO (FCbO) [33], and CbO realization InClose [4]. These algorithms have same worst-case theoretical complexity as NextClosure and CbO, but often perform better in practice. Here we present CbO algorithm, for it is both quite simple, efficient, and embodies a useful technique that can be employed for solution of other problems.

**Close-by-One (CbO) Algorithm** We introduce the following notation for CbO in the bottom-up strategy, when computation is performed by adding objects:

- $min(X)$ ($max(X)$) returns the object with the least (largest) number from set $X$;

- $(A, i)$ denotes $(A \cup \{i\})''$,

- $suc(A)$ is the set of all *children* of set $A$, i.e., concepts of the form $(A \cup \{i\})''$ such that $\min((A \cup \{i\})'' \setminus A) = i$. Pairs $(X, suc(X))$ are edges of the tree whose vertices are concepts (or their extents).

- $prev(A)$ returns *parent* (in the CbO-tree) of the concept with extent $A$.

- $nexti(A)$ returns the number of the next object $i$ for testing whether $(A, i)$ is the descendant of $A$.

**Close-by-One Algorithm**

0. $A:=\emptyset$, $nexti(A):=1$, $prev(A):=\emptyset$, $suc(A):=\emptyset$.
1.  **until** $A = \emptyset$ and $nexti(A) > |G|$ **do**
2.  **begin until** $nexti(A) > |G|$ **do**
3.      **begin** $i: = nexti(A)$
4.          **if** $min((A \cup \{i\})'' \setminus A) \geq i$ **then**
5.          **begin** $suc(A) \leftarrow address(A, i)$
6.              $prev(A, i):=A$
7.              $nexti(A):= nexti(A) + 1$
8.              $nexti((A, i)):= min (\{j \mid i < j \& j \notin (A, i)\})$
9.              $A:= (A, i)$, **output** $(A.A')$
10.         **end**
11.         **else** $nexti(A):=nexti(A) + 1$
12.     **end**
13.     $A:=prev(A)$
14. **end**

## 6.3 Canonical generation of extents

**Definition 6.1.** *We give a recurrent definition of* canonical generation *of extent.*

1. $(\emptyset)''$ *is a canonical generation of extent.*

2. $(A \cup \{i\})''$ *is canonically generated extent if $A \subseteq G$ is a canonically generated extent, $i \in G \setminus A$, and $min((A \cup \{i\})'' \setminus A) = i$.*

In terms of bracket notation canonical generation can be defined as follows: the extent generation is canonical if for every subsequence of symbols of this generation $\ldots x]Y)\ldots$ where $x \in M$ and $Y \subseteq M$, the number of every $y \in Y$ is larger than the number of $x$.

**Lemma 6.1.** Every extent $A$ has a unique canonical generation.

**Proof.** The proof is given by the following algorithm of canonical generation of extent $A$.

0. $C := \emptyset$, $i := 0$
1.  **from** $C = \emptyset$ **until** $C = A$ **do**

   2. **begin**

   3.       $i := min(A \setminus C), C := (C \cup \{i\})''$

   4. **end**

Due to duality of objects and attributes, CbO algorithm can also be executed in the top-down strategy, when attributes are added one by one, starting from the empty set of attributes. Below we present examples of CbO tree for bottom-up and top-down strategies for the context from Example 5.1.

## CbO in the bottom-up strategy (adding objects)

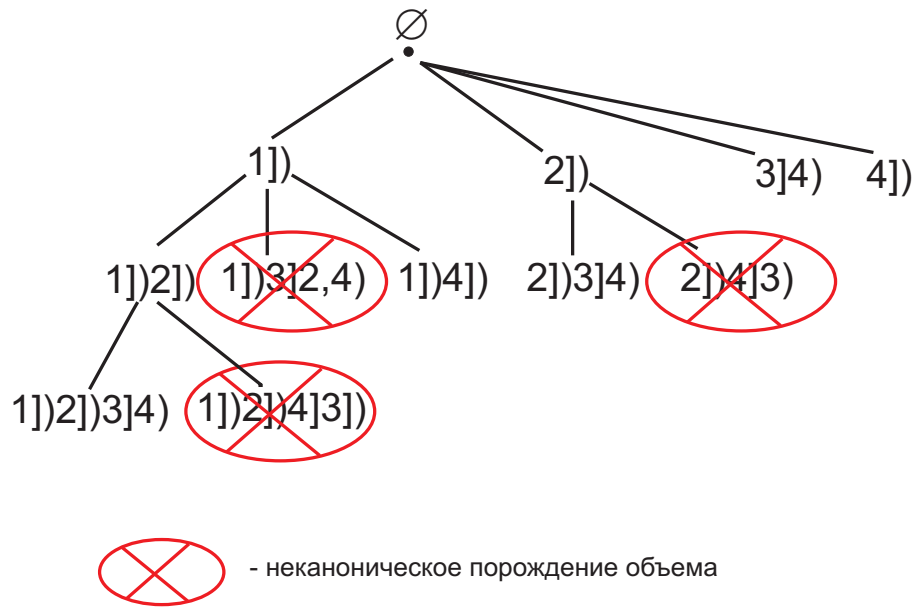

⊗ - неканоническое порождение объема

## CbO in the top-down strategy (adding attributes)

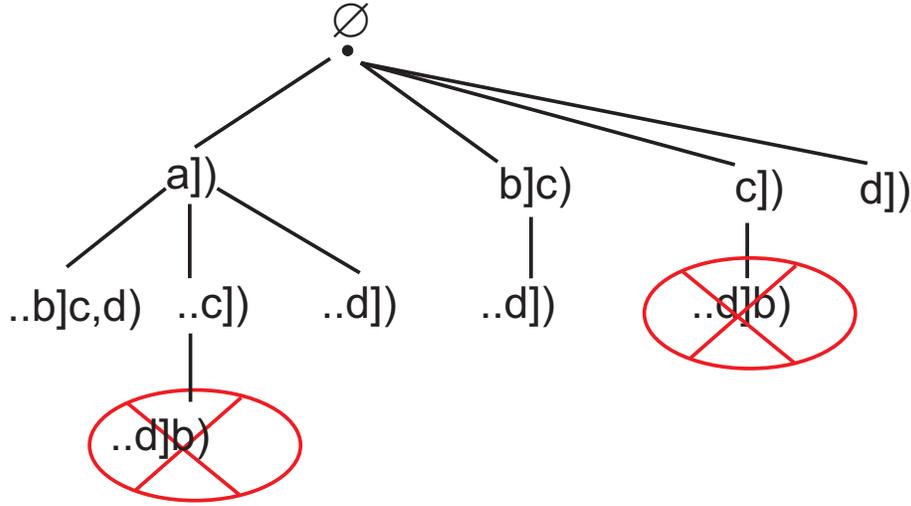

### 6.3.1  Algorithmic complexity of CbO

Consider the bottom-up approach, where one proceeds with objects. First, we estimate the size of the CbO tree. Denote by $C$ the set of all concepts. According to Lemma 6.1, the number of canonical vertices of the tree is the same as the number of all concepts, i.e., $|C|$. Every canonical vertex has at most $|G|$ children (actually, the estimate is tight only for the first level, at the $i$th level the number of children of a vertex is less or equal to $|G| - i$), so the total number of the vertices of CbO tree is $O(|C| \cdot |G|)$. At each canonical vertex one computes $(\cdot)'$ and $(\cdot)''$, which takes $O(|G| \cdot |M|)$ time if we consider that complexity of multiplying $n$-bit vectors is $O(n)$. These operations take only one bit string of size $|G|$, but one needs $O(|G| + |M|)$ space to store the new extent and intent. Hence, the total time spent by CbO is $O(|C| \cdot |G|^2 \cdot |M|)$ and total space is $O(|C| \cdot |G| \cdot (|G| + |M|))$.

Now consider the delay of CbO. The longest computation between two outputs or between the last output and termination looks as follows: in the lowest level of the tree, one backtracks several (at most $|G|$) times and after backtracking computes at most $|G|$ times operations $(\cdot)'$ and $(\cdot)''$, which takes $O(|G| \cdot |M|)$ time. Upon (at most $|G|$) unsuccessful (i.e., without producing a canonical generation) backtracks one reaches the root of the tree and tries to produce a canonical generation. If no canonical generation is generated, the algorithm terminates. So, the delay would be $O(|C| \cdot |G|^3 \cdot |M|)$ if we proceed in this way. However, a small trick can be done to improve the delay. Let us output the canonical generation not at the time of its computation, but when one backtracks from the respective vertex upwards. Then the time before the next concept output becomes smaller by factor $|G|$, i.e., $O(|G|^2 \cdot |M|)$, which is in good accordance with the total worst-case complexity of CbO estimated above.

If one computes the concepts in the dual, top-down way, by increasing the set of attributes, then the total time spent by CbO is $O(|C| \cdot |M|^2 \cdot |G|)$, the total space is $O(|C| \cdot |M| \cdot (|G| + |M|))$, and the delay is $O(|M|^2 \cdot |G|)$. Obviously, if the goal is just to compute the concepts, it is more efficient to compute them bottom-up if $|G| < |M|$ and to compute them top-down if $|M| < |G|$.

The following heuristic for speeding concept computation in practice is related to reducing the number of non-canonical generations. Before computing top-down we order objects by increasing cardinalities of object intents, i.e., if $|g'| < |h'|$, then $g$ gets a smaller number than $h$. By this reordering we diminish the probability of the fact that intersection of two object intents is contained in an object intent with smaller number (i.e., violation of canonicity), since it is not likely that intersection of big sets is contained in a small set. This heuristic does not improve the worst-case complexity of CbO, but often makes it run faster in practice.

## 6.4 Computing the covering relation of the concept lattice

CbO does not compute the covering relation of the concept lattice, which is defined on pairs of concepts and gives the graph of the diagram of the concept lattice. Note that to be a neighbor in CbO-tree is not equivalent to be a neighbor in the concept diagram. If $x$ lies below $y$ in the CbO-tree, then $x \leq y$ in the concept lattice. However, if $x \leq y$ in the concept lattice, this does not mean that the canonical generation of $x$ lies below the canonical generation of $y$ in the CbO-tree. So, the CbOtree is very weakly related to the diagram of the concept lattice and cannot be used directly for generating the covering relation. To compute the covering relation of the concept lattice when the CbO-tree is already there, one can proceed as follows. Say, in the top-down strategy, for every canonically generated concept $(A, B)$ represented by intent $B$ one first computes candidates for the lower neighbors, which are among sets of the form $(B \cup \{m\})''$, $m \in M \setminus B$, then one takes only inclusion-minimal from them. These intents correspond to concepts covered by $(A, B)$ in the concept lattice. It remains to store the addresses of these concepts in the vertex of the CbO-tree that corresponds to the canonical generation of $(A, B)$.

The complexity of the procedure is as follows. For the canonical vertex of the CbO tree that corresponds to concept $(A, B)$ computing $(A \cup \{g\})'$ and $(A \cup \{g\})''$ takes $O(|G| \cdot |M|)$ time and $O(|G|)$ space for every $g \in G \setminus A$, so computing all $|G|$ candidates for the upper neighbors of an element in the concept lattice takes $O(|G| \cdot |G| \cdot |M|)$ time and $O(|G| \cdot |G|)$ space. Then one selects minimal (w.r.t. lattice order) candidates from all $O(|G|)$ candidates, comparing all pairs of candidates, which takes $O(|G|^2 \cdot |G|)$ time and $O(|G|^2)$ space, so the total time complexity per canonical vertex is $O(|G|^2 \cdot (|G| + |M|))$ and total space complexity is $O(|G|^2)$. Therefore, the total time and space complexity for the whole algorithm are $O(|C| \cdot |G|^2 \cdot (|G| + |M|))$ and $O(|C| \cdot |G|^2)$, respectively.

## 6.5 Computing implication bases

The canonical implication base (Duquenne-Guigues base, stembase, minimal implication base) proposed in [23] was defined above in terms of pseudo-intents [19]. The first intractability result concerning Duquenne-Guigues basis is that its size can be exponential in the input size, i.e., in the size of the context. Consider the following context $K_{\exp,3}$.

| $G \setminus M$ | $m_0$ | $m_1$ | $m_2$ | $m_3$ | $m_4$ | $m_5$ | $m_6$ |
|---|---|---|---|---|---|---|---|
| $g_1$ |  |  | × | × |  | × | × |
| $g_2$ |  | × |  | × | × |  | × |
| $g_3$ |  | × | × |  | × | × |  |
| $g_4$ | × |  | × | × | × | × | × |
| $g_5$ | × | × |  | × | × | × | × |
| $g_6$ | × | × | × |  | × | × | × |
| $g_7$ | × | × | × | × |  | × | × |
| $g_8$ | × | × | × | × | × |  | × |
| $g_9$ | × | × | × | × | × | × |  |

Here set $\{m_1, m_2, m_3\}$ is a pseudo-intent, since it is not closed, but every subset of it is closed. Replacing $m_1$ with $m_4$ and/or $m_2$ with $m_5$ and/or $m_3$ with $m_6$, we also obtain pseudo-intents. So, here we have $2^3 = 8$ pseudo-intents: $\{m_1, m_2, m_3\}, \{m_1, m_2, m_6\}, \{m_1, m_5, m_3\}, \{m_1, m_5, m_6\}, \{m_4, m_2, m_3\}, \{m_4, m_2, m_6\}, \{m_4, m_5, m_3\}, \{m_4, m_5, m_6\}$.

In general, we can have the following situation

| $G \setminus M$ | $m_0$ | $m_1 \ldots m_n$ | $m_{n+1} \ldots m_{2n}$ |
|---|---|---|---|
| $g_1$ $\vdots$ $g_n$ | | $\neq$ | $\neq$ |
| $g_{n+1}$ $\vdots$ $\vdots$ $\vdots$ $g_{3n}$ | $\times$ $\vdots$ $\vdots$ $\vdots$ $\times$ | $\neq$ | |

The set $\{m_1, \ldots, m_n\}$ is a pseudo-intent. Replacing $m_i$ with $m_{n+i}$ independently for each $i$, one obtains all $2^n$ pseudo-intents.

Moreover, the problem of counting pseudo-intents is #P-hard.

**Proposition.** The following problem is #$P$-hard.

    **INPUT** A formal context $K = (G, M, I)$

    **OUTPUT** The number of pseudo-intents of $K$

**Proof**: by reduction from the problem of counting all (inclusion) minimal covers proved to be #$P$-complete in [35]. Recall that for a graph $(V, E)$ a subset $W \subseteq V$ is a *vertex cover* if every edge $e \in E$ is incident to some $w \in W$. Consider the context of the following form, where $\bar{I}$ is the complement of the edge-vertex graph incidence matrix, attributes are in one-to-one correspondence with graph vertices from set $V$ and objects are in one-to-one correspondence with edges of the graph from set $E$

| $G \setminus M$ | $m_0$ | $m_1, \ldots \ldots \ldots, m_{|V|}$ |
|---|---|---|
| $g_1$ | | |
| $\vdots$ | | |
| $\vdots$ | | $\bar{I}$ |
| $\vdots$ | | |
| $g_{|E|}$ | | |
| $g_{|E|+1}$ | $\times$ | |
| $\vdots$ | $\vdots$ | |
| $\vdots$ | $\vdots$ | $\neq$ |
| $\vdots$ | $\vdots$ | |
| $g_{|E|+|V|}$ | $\times$ | |

Every pseudo-intent of this context is in one-to-one correspondence with a minimal vertex cover of the graph. Indeed, let $W \subseteq V$ be a vertex cover of $G$, then $W$ is a pseudo-intent of the context, since $W$ is not closed and all its subsets are closed.

Most likely the problem of counting pseudo-intents does not lie in class #P, since the decision problem of determining whether an arbitrary subset of attributes is a pseudo-intent is coNP-complete [5]. Moreover, pseudo-intents cannot be generated in lexicographic order or reverse lexicographic order with polynomial delay if P$\neq$ coNP [15, 5]. The problem of computing pseudo-intents is at least as hard as the problem of enumerating all minimal transversals of a hypergraph [15].

Example with context $K_{\exp,n}$ presented above shows also that the number of minimal generators and the number of proper premises can also be exponential in the size of the context, since in this example every pseudo-intent is a minimal generator and a proper premise, i.e., a premise of the direct basis.

## 6.6  Algorithms for computing implication bases

In contrast to the situation with computing the concepts, not much algorithms for computing the canonical Duquenne-Guigues implication base are known. For long time the only algorithm used for computing the canonical base was the one based on NextClosure [17]. NextClosure computes closed sets w.r.t. an arbitrary closure operator on $M$ in the lectic order, i.e., on lexicographic order of characteristic vectors of attribute subsets. Assume that elements of $A$ are linearly ordered, e.g., numbered. Formally, $A \subseteq M$ is *lectically smaller* than $B \subseteq M$ if

$$\exists b \in B \setminus A \quad \forall a \in A \ (a < b \ \rightarrow a \in B).$$

Subset $A$ is lectically smaller than subset $B$ if it is $B$ which contains the least element where $A$ and $B$ differ. Obviously, the lectic order is a linear extension of the set-theoretic containment on subsets of $M$: if $A \subset B$, then $A$ is lectically smaller than $B$.

---

**Algorithm 1** NEXT CLOSURE$(A, M, \mathcal{L})$

---

**Input:** Closure operator $X \mapsto \mathcal{L}(X)$ on set of attributes $M$ and subset $A \subseteq M$.
**Output:** Lectically next closed set after set $A$.
    **for all** $m \in M$ in reverse order **do**
        **if** $m \in A$ **then**
            $A := A \setminus \{m\}$
        **else**
            $B := \mathcal{L}(A \cup \{m\})$
            **if** $B \setminus A$ does not contain elements $< m$ **then**
                **return** $B$
    **return** $\bot$

---

Computing Duequenne-Guigues base with NextClosure uses the fact that the union of all closed and pseudo-closed subsets of set $M$ makes a closure system (i.e., if two subsets belong to this set, then their intersection also belongs to this set). The closure operator $^{\bigcirc} : M \to M$ [23, 31] that corresponds to this closure system is given by:

$$A^+ = A \cup \bigcup \{P'' \mid P \to P'', P \text{ is a pseudo-intent}, P \subset A\}.$$

Then $A^{\bigcirc} = A^{++\ldots+}$, where $A^{\bigcirc+} = A^{\bigcirc}$. By definition, operator $(\cdot)^{\bigcirc}$ is extensive, monotone, and idempotent, i.e., it is a closure operator. By definition of pseudo-intent $A^{\bigcirc}$ is either a pseudo-intent if $A'' \neq A$ or an intent if $A'' = A$.

To compute $A^+$ one can use an arbitrary algorithm for computing the implicational closure L, which is defined in the following way:

$$\mathcal{L}(X) = \bigcap \{Y \mid X \subseteq Y \subseteq M, \forall (A \to B) \in \mathcal{L} \quad A \nsubseteq Y \text{ or } B \subseteq Y\}.$$

The simplest algorithm SimpClosure, which computes $\mathcal{L}(X)$ for a subset of attributes $X \subseteq M$, just iterates the application of implications to subset $X$:

---
**Algorithm 2** IMPLICATIONAL CLOSURE$(X, \mathcal{L})$
---
**Input:** Set of attributes $X \subseteq M$ and set $\mathcal{L}$ of implications on $M$.
**Output:** Implicational closure of $X$ w.r.t. implictions in $\mathcal{L}$.
   **repeat**
        *stable* := **true**
        **for all** $A \to B \in \mathcal{L}$ **do**
            **if** $A \subseteq X$ **then**
                $X := X \cup B$
                *stable* := **false**
                $\mathcal{L} := \mathcal{L} \setminus \{A \to B\}$
   **until** *stable*
   **return** $X$

---

In the worst case this algorithm performs a number of operations quadratic w.r.t. the number of implications in L, more precisely, $O(m \cdot n^2)$, where $|\mathcal{L}| = n, |M| = m$, since one implication can be applied at each iteration of the cycle **repeat**, whereas each cycle is performed in time $O(m)$.

Implicative closure L can be computed by means of algorithm LinClosure [7]. This algorithm is well-known in the theory of relational database (see its description in [29]), where it is used for computing the closure of attribute sets w.r.t. a set of functional dependencies.

Consider the following example. Let $M = \{a, b, c, d, e, f, g\}$, $X = \{f\} \subseteq M$, and implications $a \to b, b \to c, c \to d, d \to e, e \to f, f \to g$ be given. For implications ordered in this way SimpClosure needs to pass through the whole current list of implications until it reaches the last one. The premise of this implication matches the current subset and is applied, after which it is deleted and the number of implications decreases by one.

In contrast to SimpClosure, algorithm LinClosure uses a special data structure, which can be represented by a bipartite graph. The vertices of

---

**Algorithm 3** LINCLOSURE($X, \mathcal{L}$)

---

**Input:** Set of attributes $X \subseteq M$ and set of implications $\mathcal{L}$ on set $M$.

**Output:** Closure of set $X$ w.r.t. set of implications $\mathcal{L}$.

    **for all** $A \rightarrow B \in \mathcal{L}$ **do**
        $count[A \rightarrow B] := |A|$
        **if** $|A| = 0$ **then**
            $X := X \cup B$ структуру
        **for all** $a \in A$ **do**
            добавить $A \rightarrow B$ в $list[a]$
    $update := X$
    **while** $update \neq \varnothing$ **do**
        выбрать $m \in update$
        $update := update \setminus \{m\}$
        **for all** $A \rightarrow B \in list[m]$ **do**
            $count[A \rightarrow B] = count[A \rightarrow B] - 1$
            **if** $count[A \rightarrow B] = 0$ **then**
                $add := B \setminus X$
                $X := X \cup add$
                $update := update \cup add$
    **return** $X$

---

one part correspond to attributes from $M$ and the vertices of the second part correspond to implications supplied with counters, which are denoted by $count[A \rightarrow B]$. There is an edge between vertex staying for $a \in M$ and vertex staying for implication $A \rightarrow B$ if $a \in A$.

LinClosure runs by passing through all edges of this bipartite graph. When an edge is passed, it is deleted from the list and the counter of the vertex staying for the respective implication is decreased by one. When $count[A \rightarrow B]$ becomes equal to zero (and the respective vertex has no more incident edges), then the current set of attributes, initiated with $X$, is updated to include $B$. Since the number of edges in the bipartite graph is equal to the sum of sizes of all implications, LinClosure has time complexity linear w.r.t. this sum complexity, or more precise, $O(m \cdot n)$. Theoretically, this is a better complexity than that of SimClosure, however, due to the more complex data structure, LinClosure can be slower than SimClosure in practice.

The following algorithm from [31] computes closed sets of attributes w.r.t. closure operator L (which are intents or pseudo-intents in the original context) and checks whether the subset is closed or not. If the subset is not closed, then it is a pseudo-intent.

---

**Algorithm 4** DUQUENNE-GUIGUES BASE($M, ''$)

---

**Input:** Closure operator $X \mapsto X''$ on $M$, e.g., given by formal context $(G, M, I)$.

**Output:** Duquenne-Guigues basis for the closure operator.

   $\mathcal{L} := \varnothing$

   $A := \varnothing$

   **while** $A \neq M$ **do**

      **if** $A \neq A''$ **then**

         $\mathcal{L} := \mathcal{L} \cup \{A \to A''\}$

      $A :=$ NEXT CLOSURE($A, M, \mathcal{L}$)

   **return** $\mathcal{L}$

---

The optimized version of this algorithm proposed in [31] uses the following facts:

Let $i$ be the maximal element of $A$ and $j$ is the minimal element of $A'' \setminus A$. Consider the following two cases:

$j < i$: While $m > i$, set $\mathcal{L}(\mathcal{A}^\bigcirc \cup \{\})$) will be rejected by NextClosure, since it contains $j$. Hence, it is worth to skip all $m > i$ and continue as if $A^\bigcirc$ was rejected by NextClosure.

$i < j$: In this case one can show that the lectic next intent or pseudo-intent after $A^\bigcirc$ is $A''$. Therefore, $A''$ can be used at the next step instead of $A^\bigcirc$.

These facts can be used in the following algorithm from [31].

Since NextClosure and its modifications generate both intents and pseudo-intents, there were attempts to construct more efficient algorithms that generate only pseudo-intents. An incremental algorithm of this kind, which according to experimental results is quite fast, was proposed in [31]. The general idea of the algorithm consists in distinguishing two types of implications in the context: $x$-modified and $x$-stable. When a new attribute $x$ arrives implication $A \to B$ is $x$-modified if $A \setminus \{x\} \to x$ and it is $x$-stable otherwise. Being faster in practice, this algorithm has the same theoretical worst-case complexity as NextClosure.

---

**Algorithm 5** Duquenne-Guigues basis $(M, '')$, optimized version

---

**Input:** Closure operator $X \mapsto X''$ on $M$, e.g., given by formal context $(G, M, I)$.

**Output:** Duquenne-Guigues basis for the closure operator.

   $\mathcal{L} := \varnothing$

   $A := \varnothing$

   $m :=$ the least element of $M$

   **while** $A \neq M$ **do**

       **if** $A \neq A''$ **then**

          $\mathcal{L} := \mathcal{L} \cup \{A \to A''\}$

       **if** $A'' \setminus A$ does not contain elements $< m$ **then**

          $A := A''$

          $m :=$ the largest element of $M$

       **else**

          $A := \{a \in A \mid a \leq m\}$

       **for all** $l \leq m \in M$ in reverse order **do**

          **if** $l \in A$ **then**

             $A := A \setminus \{l\}$

          **else**

             $B := \mathcal{L}(A \cup \{l\})$

             **if** $B \setminus A$ does not contain elements $< l$ **then**

                $A := B$

                $m := l$

                **exit for**

   **return** $\mathcal{L}$

---

**Computing direct basis**. The premises of implications of the direct basis are proper premises. We present the following result from [19] showing that the computation of proper premises can be reduced to the famous problem of generating minimal transversal of a hypergraph.

Let $g \downarrow m$ denote that $m \notin g'$, but $m \in h'$ for any $g, h \in G$ such that $g' \subset h'$, so $g$ does not have $m$, but any object $h$ with object intent larger than that of $g$ will have $m$. Let $A^{\bigcirc} = A'' \setminus (A \cup \bigcup_{n \in A}(A \setminus \{n\}))$, so $A^{\bigcirc}$ is what cannot be derived from proper subsets of $A$.

**Proposition 6.1.** [19] $P$ is a proper premise with $m \in P^{\bigcirc}$ iff $P$ is minimal

w.r.t. the property that

$$(M \setminus g') \cap P \neq \emptyset$$

holds for all $g \in G$ such that $g \downarrow m$.

The proof of the Proposition 6.1 is based on the simple fact that if $P$ is a proper premise of $m$, then it is not contained in any object intent not having $m$, hence it should have at least one attribute in common with the complement of every such object intent.

The proposition implies that all proper premises implying $m$ are found as minimal transversals of the hypergraph formed by complements of the object intents not having $m$. Generating minimal hypergraph transversals is a famous problem in computer science, see review [16] for the complexity analysis and efficient algorithms.

Having the direct basis, one can obtain the canonical basis, e.g., by an effcient algorithm (with quadratic worst-case time complexity) from the database theory [7, 29] for minimizing the set of functional dependencies.

# Exercises

| Числа | Составное | Чётное | Нечётное | Простое | Квадрат |
|-------|-----------|--------|----------|---------|---------|
| 1     |           |        | ×        |         | ×       |
| 2     |           | ×      |          | ×       |         |
| 3     |           |        | ×        | ×       |         |
| 4     | ×         | ×      |          |         | ×       |
| 5     |           |        | ×        | ×       |         |
| 6     | ×         | ×      |          |         |         |
| 7     |           |        | ×        | ×       |         |
| 8     | ×         | ×      |          |         |         |
| 9     | ×         |        | ×        |         | ×       |
| 10    | ×         | ×      |          |         |         |

1. Prove that the union of the set of intents and pseudo-intents is a closure system w.r.t. set-theoretic intersection.

2. What is the worst-case complexity of the algorithm for computing canonical generation of an extent (Algorithm ??).

3. What is the worst-case complexity of reordering objects w.r.t. the cardinalities of object intents?

4. For the context in Example ??  compute Duquenne-Guigues basis with NextClosure and optimized algorithm, compare the number of computation steps.

5. Determine worst-case time and space complexity of NextClosure and the optimized algorithm.

# Index